\documentclass[twocolumn,tighten,times]{aastex63} %,twocolappendix ,linenumbers

\usepackage{amsmath} %%-->for use of eqref
\usepackage{amssymb}
\received{\today}
\revised{\today}
\accepted{\today}
\newcommand{\galprop}{\textsc{GalProp}}
\newcommand{\helmod}{\textsc{HelMod}}

\shorttitle{Fluorine Excess in Cosmic Rays
}
\shortauthors{Boschini et al.}

\begin{document}

\title{
A Hint of a Low-Energy Excess in Cosmic-Ray Fluorine
}

\author[0000-0002-6401-0457]{M.~J.~Boschini}
\affiliation{INFN, Milano-Bicocca, Milano, Italy}
\affiliation{CINECA, Segrate, Milano, Italy}

\author[0000-0002-7669-0859]{S.~{Della~Torre}}
\affiliation{INFN, Milano-Bicocca, Milano, Italy}

\author[0000-0003-3884-0905]{M.~Gervasi}
\affiliation{INFN, Milano-Bicocca, Milano, Italy}
\affiliation{Physics Department, University of Milano-Bicocca, Milano, Italy}

\author[0000-0003-1942-8587]{D.~Grandi}
\affiliation{INFN, Milano-Bicocca, Milano, Italy}
\affiliation{Physics Department, University of Milano-Bicocca, Milano, Italy}

\author[0000-0003-1458-7036]{G.~J\'{o}hannesson} 
\affiliation{Science Institute, University of Iceland, Dunhaga 3, IS-107 Reykjavik, Iceland}
\affiliation{NORDITA,  Roslagstullsbacken 23, 106 91 Stockholm, Sweden}

\author[0000-0002-2168-9447]{G.~{La~Vacca}}
\affiliation{INFN, Milano-Bicocca, Milano, Italy}
\affiliation{Physics Department, University of Milano-Bicocca, Milano, Italy}

\author[0000-0002-3729-7608]{N.~Masi}
\affiliation{INFN, Bologna, Italy}
\affiliation{Physics Department, University of Bologna, Bologna, Italy}

\author[0000-0001-6141-458X]{I.~V.~Moskalenko} 
\affiliation{Hansen Experimental Physics Laboratory, Stanford University, Stanford, CA 94305}
\affiliation{Kavli Institute for Particle Astrophysics and Cosmology, Stanford University, Stanford, CA 94305}

\author{S.~Pensotti}
\affiliation{INFN, Milano-Bicocca, Milano, Italy}
\affiliation{Physics Department, University of Milano-Bicocca, Milano, Italy}

\author[0000-0002-2621-4440]{T.~A.~Porter} 
\affiliation{Hansen Experimental Physics Laboratory, Stanford University, Stanford, CA 94305}
\affiliation{Kavli Institute for Particle Astrophysics and Cosmology, Stanford University, Stanford, CA 94305}

\author{L.~Quadrani}
\affiliation{INFN, Bologna, Italy}
\affiliation{Physics Department, University of Bologna, Bologna, Italy}

\author[0000-0002-1990-4283]{P.~G.~Rancoita}
\affiliation{INFN, Milano-Bicocca, Milano, Italy}

\author[0000-0002-7378-6353]{D.~Rozza}
\affiliation{INFN, Milano-Bicocca, Milano, Italy}
%\affiliation{Physics Department, University of Milano-Bicocca, Milano, Italy}

\author[0000-0002-9344-6305]{M.~Tacconi}
\affiliation{INFN, Milano-Bicocca, Milano, Italy}
\affiliation{Physics Department, University of Milano-Bicocca, Milano, Italy}
%\affil{institute}
%\email{\myemail}

%% Notice that each of these authors has alternate affiliations, which
%% are identified by the \altaffilmark after each name.  Specify alternate
%% affiliation information with \altaffiltext, with one command per each
%% affiliation.

%% Mark off your abstract in the ``abstract'' environment. In the manuscript
%% style, abstract will output a Received/Accepted line after the
%% title and affiliation information. No date will appear since the author
%% does not have this information. The dates will be filled in by the
%% editorial office after submission.

%$_{\phantom{1}9}^{19}$F 

\begin{abstract}

Since its launch, the Alpha Magnetic Spectrometer--02 (AMS-02) has delivered outstanding quality measurements of the spectra of cosmic-ray (CR) species, $\bar{p}$, $e^{\pm}$, and nuclei (H--O, Ne, Mg, Si, Fe), which resulted in a number of breakthroughs. The most recent AMS-02 result is the measurement of the spectrum of CR fluorine up to $\sim$2 TV. Given its very low solar system abundance, fluorine in CRs is thought to be mostly secondary, produced in fragmentations of heavier species, predominantly Ne, Mg, and Si. Similar to the best-measured secondary-to-primary boron to carbon nuclei ratio that is widely used to study the origin and propagation of CR species, the precise fluorine data would allow the origin of Si-group nuclei to be studied independently. Meanwhile, the secondary origin of CR fluorine has never been tested in a wide energy range due to the lack of accurate CR data. In this paper, we use the first ever precise measurements of the fluorine spectrum by AMS-02 together with ACE-CRIS and Voyager 1 data to actually test this paradigm. Our detailed modeling shows an excess below 10 GV in the fluorine spectrum that may hint at a primary fluorine component. We also provide an updated local interstellar spectrum (LIS) of fluorine in the rigidity range from few MV to $\sim$2 TV. Our calculations employ the self-consistent \galprop{}--\helmod{} framework that has proved to be a reliable tool in deriving the LIS of CR $\bar{p}$, $e^{-}$, and nuclei $Z\le28$.
\end{abstract}

%% Keywords should appear after the \end{abstract} command. The uncommented
%% example has been keyed in ApJ style. See the instructions to authors
%% for the journal to which you are submitting your paper to determine
%% what keyword punctuation is appropriate.


\section{Introduction} \label{Intro}
%%%%%%%%%%%%%%%%%%%%%%%%%%%%%%%%%%%%%%%%%%%%%%%%%%%
%%%%%%%%%%%%%%%%%%%%%%%%%%%%%%%%%%%%%%%%%%%%%%%%%%%

The precise data delivered by the new generation of space instrumentation allow the stellar nucleosynthesis, properties of the interstellar medium (ISM), and the origin of CRs to be probed to much finer details than was possible just a decade ago. The measurements provided by individual spacecraft, can be combined to cover the enormous range of rigidities, from few MV to tens of TV, where the individual spectra of CR species are shaped by many different processes. Their analysis and interpretation enables the discoveries of new phenomena far outside of the local solar neighborhood.

The spectrum of CR fluorine from AMS-02 \citep{2021PhRvL.126h1102A} is the latest\footnote{Spectra of Na and Al have just been published \citep{2021PhRvL.127b1101A}.} in a series of publications of spectra of most abundant species, H--O, Ne, Mg, Si, Fe \citep{2014PhRvL.113v1102A, 2015PhRvL.114q1103A, 2015PhRvL.115u1101A, 2016PhRvL.117i1103A, 2016PhRvL.117w1102A, 2017PhRvL.119y1101A, 2018PhRvL.120b1101A, 2018PhRvL.121e1103A, 2019PhRvL.122d1102A, 2019PhRvL.122j1101A, 2020PhRvL.124u1102A, 2021PhRvL.126d1104A}. Fluorine is less studied because its CR abundance is 100 and 10 times lower than that of its neighbors, oxygen and neon, correspondingly. Previous measurement of its spectrum in the range from 0.62--35 GeV nucleon$^{-1}$ was made by the HEAO-3-C2 instrument in 1979-1981 \citep{1990A&A...233...96E}. Although an excellent instrument for its epoch, HEAO-3-C2 measurements exhibit significant systematic uncertainties when compared to the AMS-02 data \citep[see detailed analyses in][]{2020ApJS..250...27B, 2021ApJ...913....5B}.

The secondary-to-primary nuclei ratio in CRs is widely used in astrophysics to derive the parameters of Galactic CR propagation for an assumed model phenomenology. The derived parameters are then applied to all CR species from electrons and antiprotons, all the way up to the Fe-group nuclei. The best measured B/C ratio is used most often. Other nuclei that are rare in the solar system, such as Li, Be, F, Sc, and V, are thought to be almost entirely produced by the fragmentation of heavier CRs species. Their precise measurements can be used to probe the origin and propagation of specific groups of nuclei: the B/C ratio for the C-N-O group, F/Si for the Si-group, and (Sc+V)/Fe for the Fe-group. However, this common wisdom has never been tested in a wide energy range due to the lack of accurate data. The first ever precisely measured fluorine spectrum by AMS-02 \citep{2021PhRvL.126h1102A} together with ACE-CRIS \citep{2013ApJ...770..117L} and Voyager 1 \citep{2016ApJ...831...18C} data offer a possibility to actually test this paradigm.

In this paper we analyze the new CR fluorine data and test their consistency with measurements of other species. We show that the spectrum of fluorine exhibits an excess below 10 GV and argue that it could be a signature of a primary fluorine component. We also provide an updated fluorine LIS in the rigidity range from few MV to $\sim$2 TV. Our calculations and interpretation employ the \galprop{}\footnote{Available from http://galprop.stanford.edu \label{galprop-site}}--\helmod{}\footnote{http://www.helmod.org/ \label{helmod-footnote}} framework that is proved to be a reliable tool in deriving the LIS of CR species \citep{2019HelMod, 2020ApJS..250...27B}.

This is the third in a series of papers where the detailed analysis of precise AMS-02 measurements taken together with Voyager 1 and ACE-CRIS data yields unexpected excesses in the spectra of CR species. The first two was an evidence of the primary lithium in CRs \citep{2020ApJ...889..167B}, and a discovery of the low-energy excess in iron \citep{2021ApJ...913....5B}.

\section{Calculations} \label{calcs}
%%%%%%%%%%%%%%%%%%%%%%%%%%%%%%%%%%%%%%%%%%%%%%%%%%%
%%%%%%%%%%%%%%%%%%%%%%%%%%%%%%%%%%%%%%%%%%%%%%%%%%%

In this work we are using the same CR propagation model with distributed reacceleration and convection that was used in our previous analyses \citep[for more details see][]{2017ApJ...840..115B,  2018ApJ...854...94B, 2018ApJ...858...61B, 2020ApJS..250...27B, 2020ApJ...889..167B, 2021ApJ...913....5B}. The latest version of the \galprop{} code for Galactic propagation of CRs is described in detail in a recent paper by \citet{2020ApJS..250...27B}, see also references therein.

Full details of the latest \helmod{} code version 4 for heliospheric propagation are provided in \citet{2019AdSpR..64.2459B}. It solves the Fokker-Planck equation for heliospheric propagation in Kolmogorov formulation backward in time \citep{2016JGRA..121.3920B}. The accuracy of the solution was tested using the Crank-Nicholson technique and found to be less than 0.5\% at low rigidities. The large number of simulated events ensures that the statistical errors are negligible compared to the other modeling uncertainties.

When comparing our calculations with data collected over extended period of time (AMS-02), variations in the solar activity are addressed in the following way. The propagation equation is solved for each Carrington rotation, and the numerical results are then combined accordingly to the AMS-02 exposure and the time period. This approach is equivalent to application of a weighted average that accounts for both exposure time and absolute counting rate variations.

The values of propagation parameters in the ISM along with their confidence limits are derived from the best available CR data using the Markov Chain Monte Carlo (MCMC) routine. Here we use the same method as described in \citet{2017ApJ...840..115B}. Five main propagation parameters, that affect the overall shape of CR spectra, were left free in the scan using \galprop{} running in 2D mode: the Galactic halo half-width $z_h$, the normalization of the diffusion coefficient $D_0$ at the reference rigidity $R=4$ GV and the index of its rigidity dependence $\delta$, the Alfv\'en velocity $V_{\rm Alf}$, and the gradient of the convection velocity $dV_{\rm conv}/dz$ ($V_{\rm conv}=0$ in the plane, $z=0$). Their best-fit values tuned to the AMS-02 data are listed in Table~\ref{tbl-prop} and are the same as obtained in \citet{2020ApJS..250...27B}. The radial size of the Galaxy does not significantly affect the values of propagation parameters and was set to 20 kpc. We also introduced a factor $\beta^\eta$ in the diffusion coefficient, where $\beta=v/c$, and $\eta$ was left free. The best fit value of $\eta=0.70$ improves the agreement at low energies, and slightly affects the choice of injection indices ${\gamma}_0$ an ${\gamma}_1$. A detailed discussion of the injection ({\it I}) and propagation ({\it P}) scenarios of the 350 GV break can be found in the works by \citet{2012ApJ...752...68V} and \citet{2020ApJS..250...27B}.

The corresponding B/C ratio also remains the same \citep[see Fig.~4 of][]{2020ApJS..250...27B}, and compares well with all available measurements: Voyager~1 \citep{2016ApJ...831...18C}, ACE-CRIS\footnote{http://www.srl.caltech.edu/ACE/ASC/level2/cris\_l2desc.html} \citep{2013ApJ...770..117L}, AMS-02 \citep{2018PhRvL.120b1101A}, ATIC-2 \citep{2009BRASP..73..564P}, CREAM \citep{2008APh....30..133A, 2009ApJ...707..593A}, and NUCLEON \citep{2019AdSpR..64.2559G}.

\begin{deluxetable}{rlcc}[tb!]
	\def\arraystretch{0.9}
	\tablewidth{0mm}
	\tablecaption{Best-fit propagation parameters for {\it I}- and {\it P}-scenarios\label{tbl-prop}}
	\tablehead{
		\colhead{Parameter}& \multicolumn{1}{l}{Units}& \colhead{Best Value}& \colhead{Error} 
	}
	\startdata
	$z_h$ & kpc &4.0 &0.6\\
	$D_0 (R= 4\ {\rm GV})$ & cm$^{2}$ s$^{-1}$  & $4.3\times10^{28}$ &0.7\\
	$\delta$\tablenotemark{a} & &0.415 &0.025\\
	$V_{\rm Alf}$ & km s$^{-1}$ &30 &3\\
	$dV_{\rm conv}/dz$ & km s$^{-1}$ kpc$^{-1}$ & 9.8 &0.8
	\enddata
	\tablenotetext{a}{The {\it P}-scenario assumes a break in the diffusion coefficient with index $\delta_1=\delta$ below the break and index $\delta_2=0.15\pm 0.03$ above the break at $R=370\pm 25$ GV \citep[for details see][]{2020ApJ...889..167B}.}
\end{deluxetable}

\section{Results and Discussion} \label{results}
%%%%%%%%%%%%%%%%%%%%%%%%%%%%%%%%%%%%%%%%%%%%%%%%%%%
%%%%%%%%%%%%%%%%%%%%%%%%%%%%%%%%%%%%%%%%%%%%%%%%%%%

\begin{figure*}[p!]%[tbh!]
	\centering
	\includegraphics[width=0.49\textwidth]{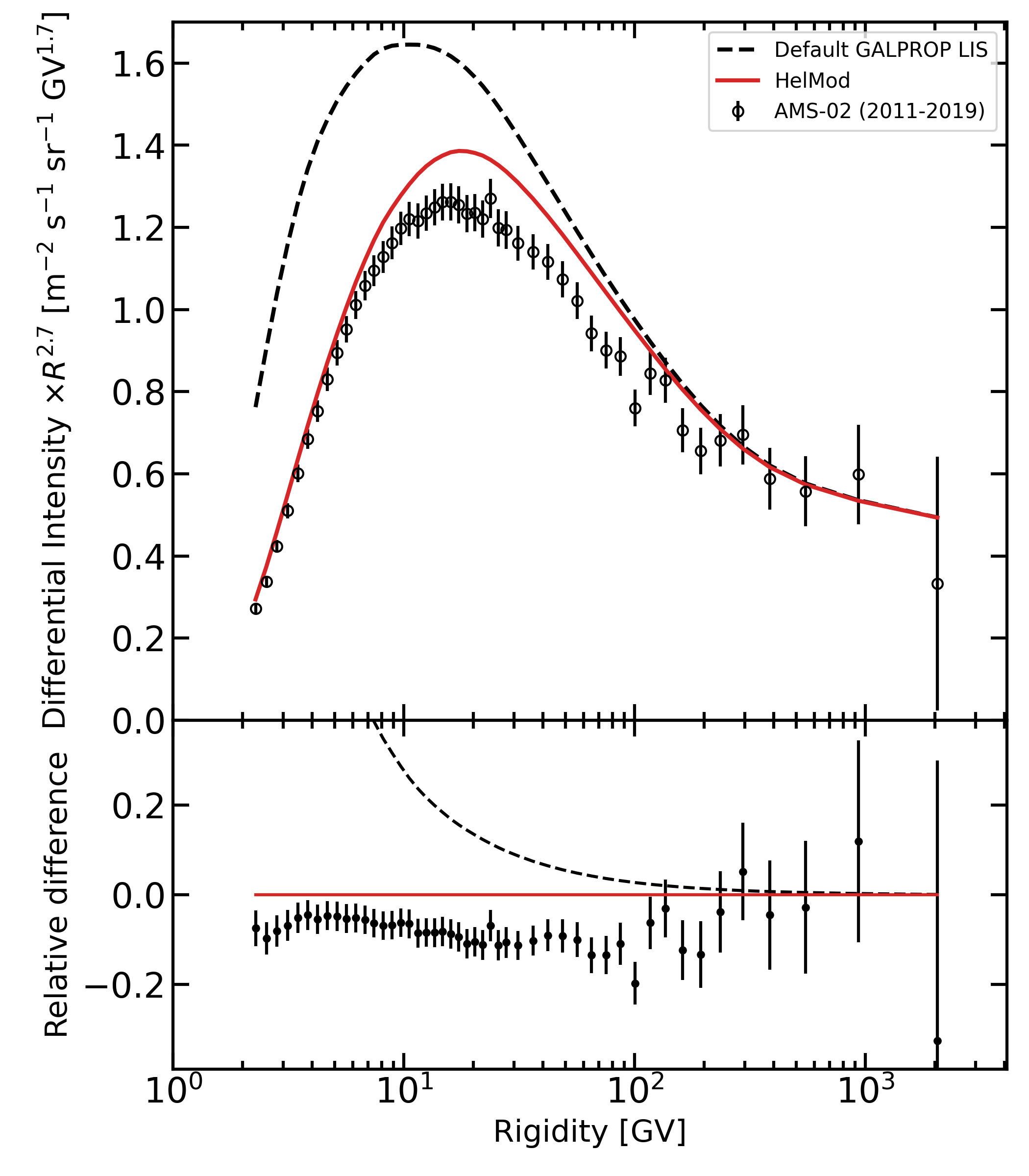}\hfill 
	\includegraphics[width=0.49\textwidth]{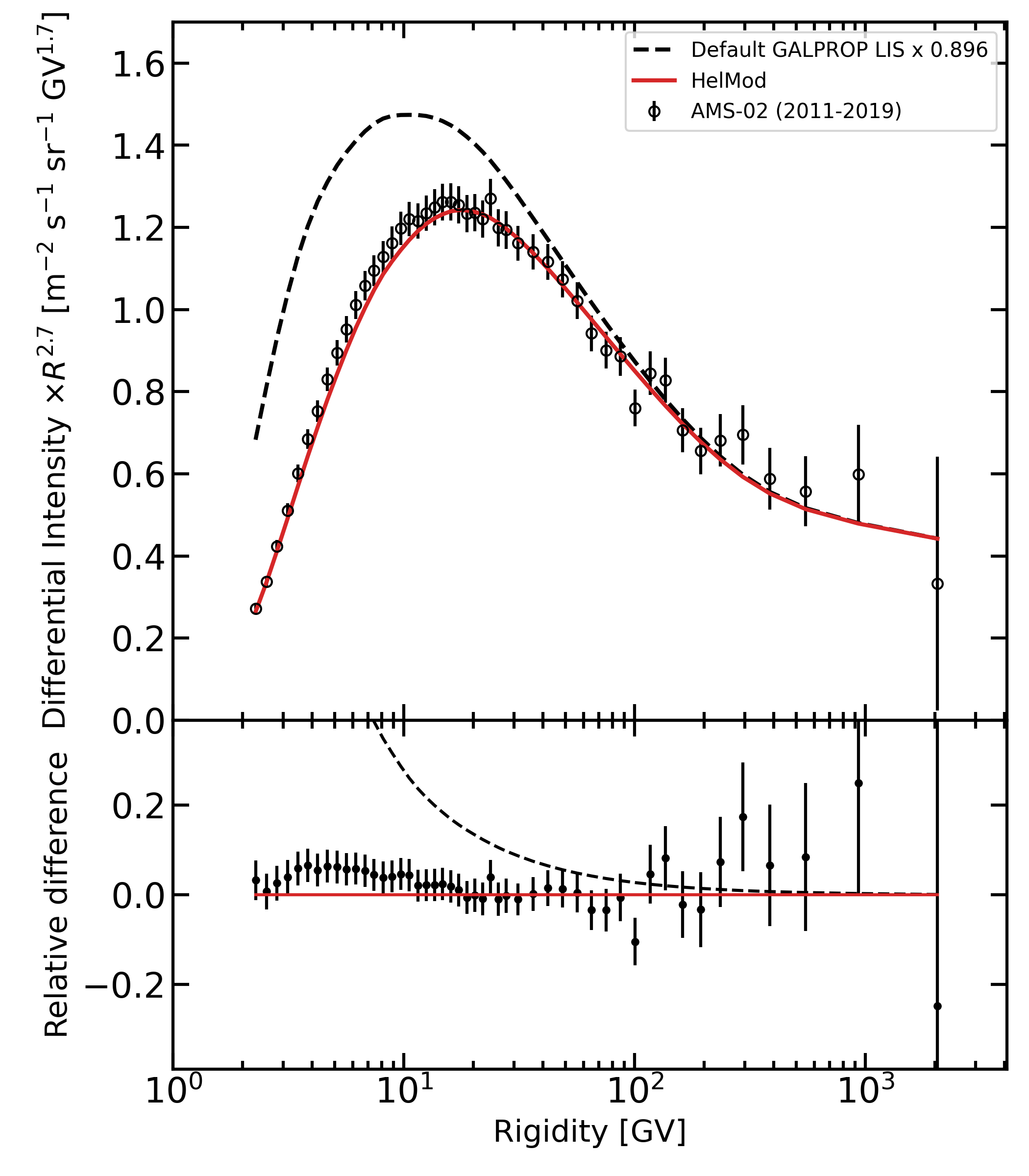}
	\includegraphics[width=0.49\textwidth]{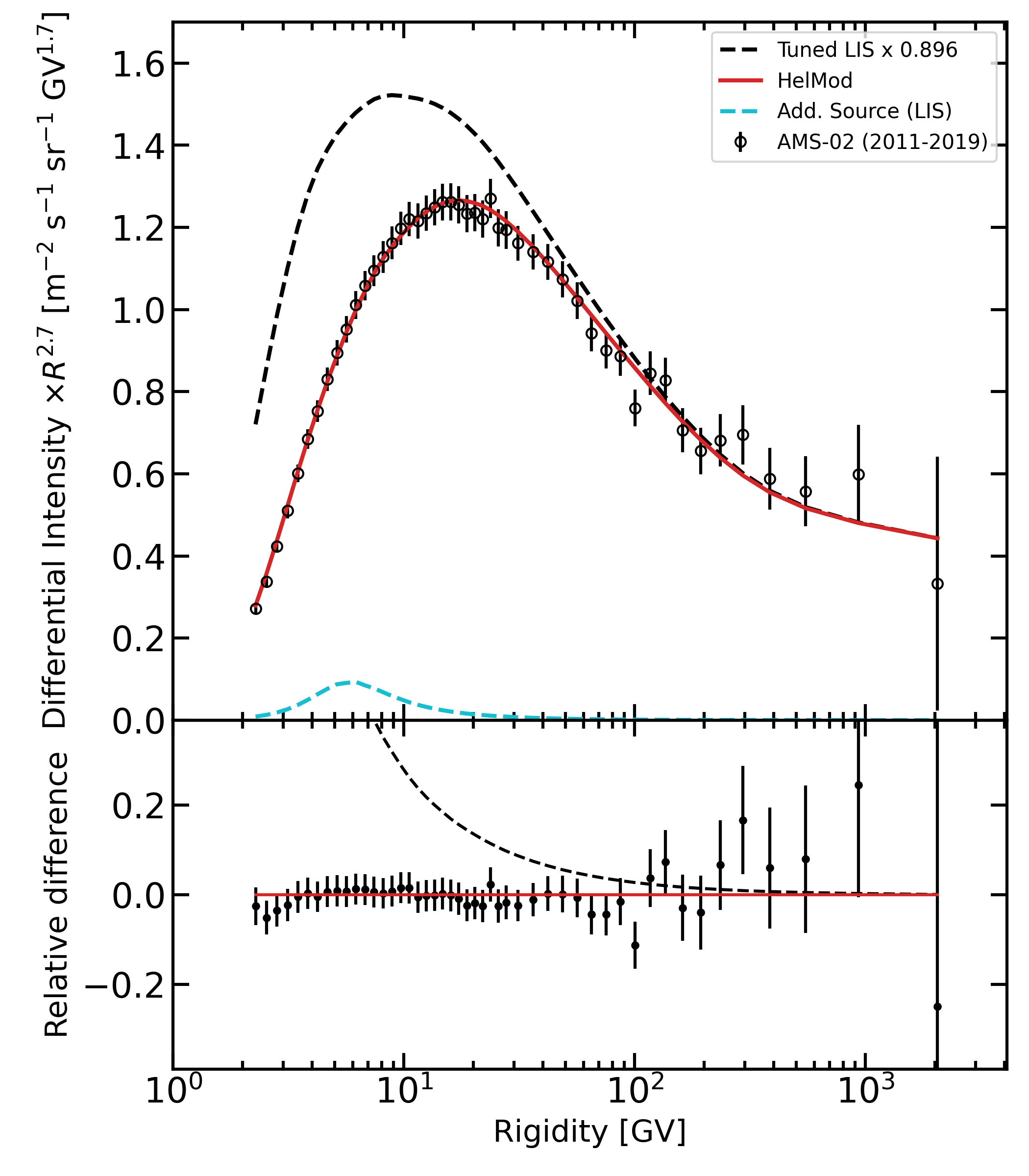} 
	\caption{
{\it Top left:} A comparison of the calculated default spectrum of secondary fluorine with the AMS-02 data \citep{2021PhRvL.126h1102A}. {\it Top right:} The calculated default spectrum is multiplied by an energy-independent factor 0.896, as discussed in the text.  {\it Bottom:} The same as in the top right, but with added primary fluorine component (see the injection spectrum in Table~\ref{tbl-inject}), where the total calculated spectrum is tuned to the data. The dashed cyan line shows separately the propagated primary fluorine component. In all panels, the dashed gray line shows the LIS, and the solid red line is the corresponding modulated spectrum. Each panel also shows the relative difference between our calculations and the data set.
	}
	\label{fig:F-spec}
\end{figure*}

\begin{figure}[tbh!]
	\centering
	\includegraphics[width=0.47\textwidth]{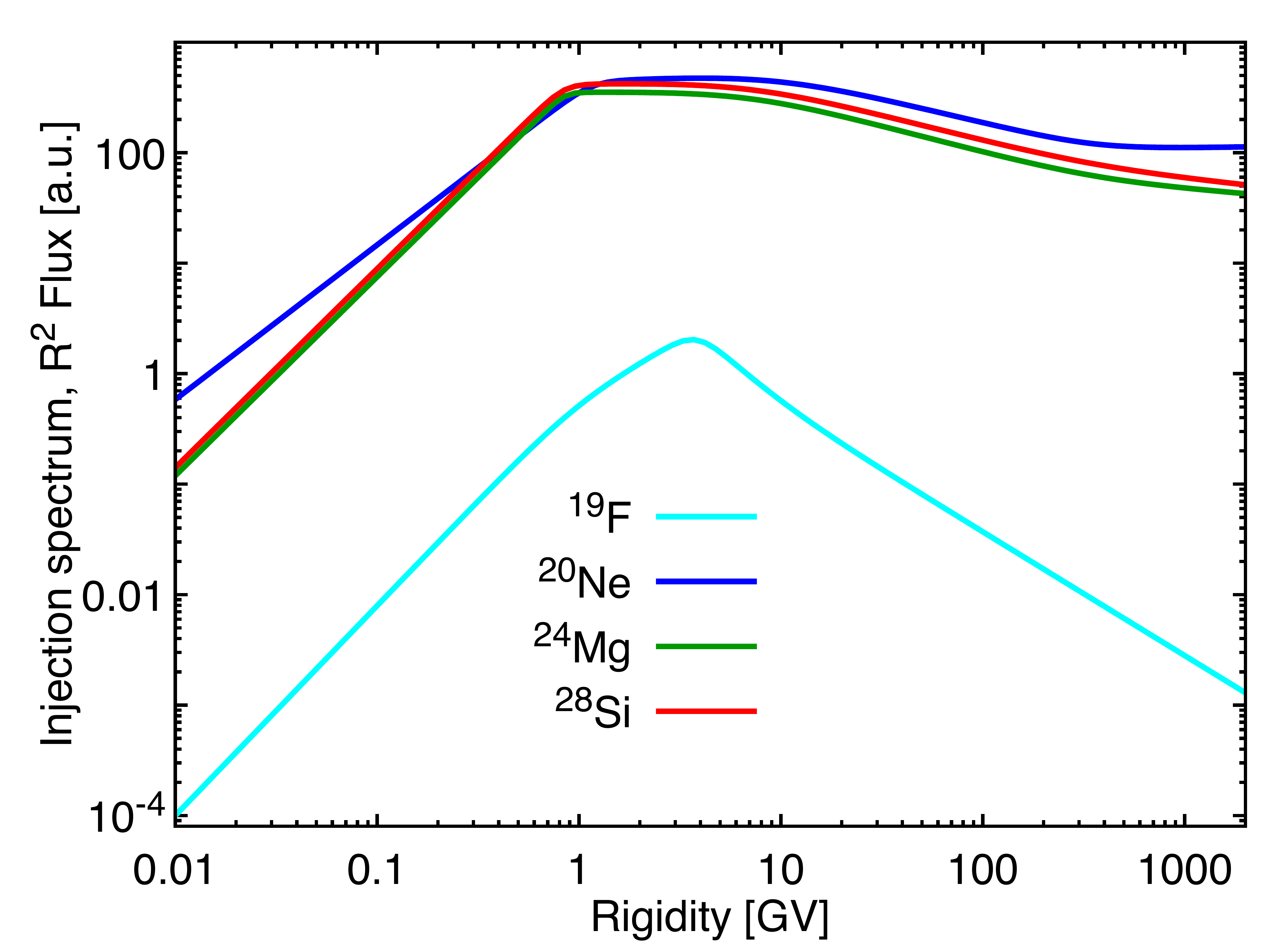}
	\caption{
The injection spectra of primary $^{19}$F and major progenitors of secondary $^{19}$F: $^{20}$Ne, $^{24}$Mg, and $^{28}$Si. The parameters for $^{19}$F are given in Table~\ref{tbl-inject}, and for $^{20}$Ne, $^{24}$Mg, and $^{28}$Si---are taken from Tables 2 and 3 in \citet{2020ApJS..250...27B}.
	}
	\label{fig:injection}
\end{figure}

Fig.~\ref{fig:F-spec} (top left panel) shows a comparison of the AMS-02 fluorine data \citep{2021PhRvL.126h1102A} with the calculated spectrum assuming the species is entirely secondary. The calculated spectral shape resembles the data, while the default normalization appears to be a bit too high over the whole range. The top right panel shows the default fluorine spectrum multiplied by a factor of 0.896 and the corresponding residuals. Here we normalized our calculated spectrum to the data in the middle rigidity range 10--100 GV. We attribute this $\sim$10\% discrepancy to the errors in the isotopic production cross sections that is well within their typical uncertainties \citep[see a detailed discussion in Appendix C.1 in][]{2020ApJS..250...27B}. The quality of the fit improves with this modification. But after the renormalization, there is now a $\sim$6--7\% excess below $\sim$10 GV.
At higher rigidities $>$100 GV, the residuals become larger, as do the error bars, that keep the overall agreement to better than 1$\sigma$. 

The bottom panel in Fig.~\ref{fig:F-spec} shows the renormalized fluorine ($\times$0.896) with the primary propagated fluorine component, where the total calculated spectrum is tuned to the data. The dashed cyan line shows separately the propagated primary fluorine component. An addition of the primary fluorine removes the excess below 10 GV and preserves the good agreement with data above $\sim$10 GV. The parameters of the injection spectrum of primary fluorine are summarized in Table~\ref{tbl-inject}. The injection spectrum is steep with index 2.90 above 4.50 GV (Fig.~\ref{fig:injection}) implying that it might be accelerated in a weak shock.

The rationale behind this spectral renormalization is simple. The effects of the nuclear structure are important below $\sim$1 GeV nucleon$^{-1}$, it is where the characteristic features in the reaction cross sections are typically observed. Above this energy the reaction cross sections are usually flat. The AMS-02 data in the heliosphere are taken above 2 GV or $\sim$1 GeV nucleon$^{-1}$, which corresponds to $\sim$2.5--3 GV or $\sim$1.25--1.5 GeV nucleon$^{-1}$ in the ISM. Therefore, the application of a single renormalization factor that takes into account possible errors in the isotopic production cross sections is well justified.

\begin{deluxetable*}{c crcl rcl rcl r}[tb!]        
\tabletypesize{\footnotesize}
        \def\arraystretch{1.1}
        \tablecolumns{12}
        \tablewidth{0mm}
        \tablecaption{The injection spectrum of primary fluorine \label{tbl-inject}}
        \tablehead{
                \multicolumn{1}{l}{}& \multicolumn{1}{c}{Source} &
                \multicolumn{10}{c}{Spectral parameters}\\
                \cline{3-12}
                \multicolumn{1}{l}{} & \multicolumn{1}{l}{abundance} & 
                \multicolumn{1}{c}{$\gamma_0$} & \multicolumn{1}{c}{$R_0$ (GV)} & \multicolumn{1}{l}{$s_0$} &
                \multicolumn{1}{c}{$\gamma_1$} & \multicolumn{1}{c}{$R_1$ (GV)} & \multicolumn{1}{l}{$s_1$} &  
                \multicolumn{1}{c}{$\gamma_2$} & \multicolumn{1}{c}{$R_2$ (GV)} & \multicolumn{1}{l}{$s_2$} &  
                \multicolumn{1}{c}{$\gamma_3$} 
                 }
	\startdata
$_{\phn9}^{19}$F & 0.63 & 0.10 & 1.00 & 0.30 & 1.00 & 3.80 & 0.32 & 3.8 & 8.0 & 0.35 & 3.12
	\enddata
  \tablecomments{See Eq.~(2) in \citet{2020ApJS..250...27B} for parameter definitions. Shown are $|s_i|$ values, note that $s_i$ is negative/positive for $|\gamma_i |\lessgtr |\gamma_{i+1} |$.}
\end{deluxetable*}

\begin{figure*}[tp!]%[tb!]
	\centering
	\includegraphics[width=0.49\textwidth]{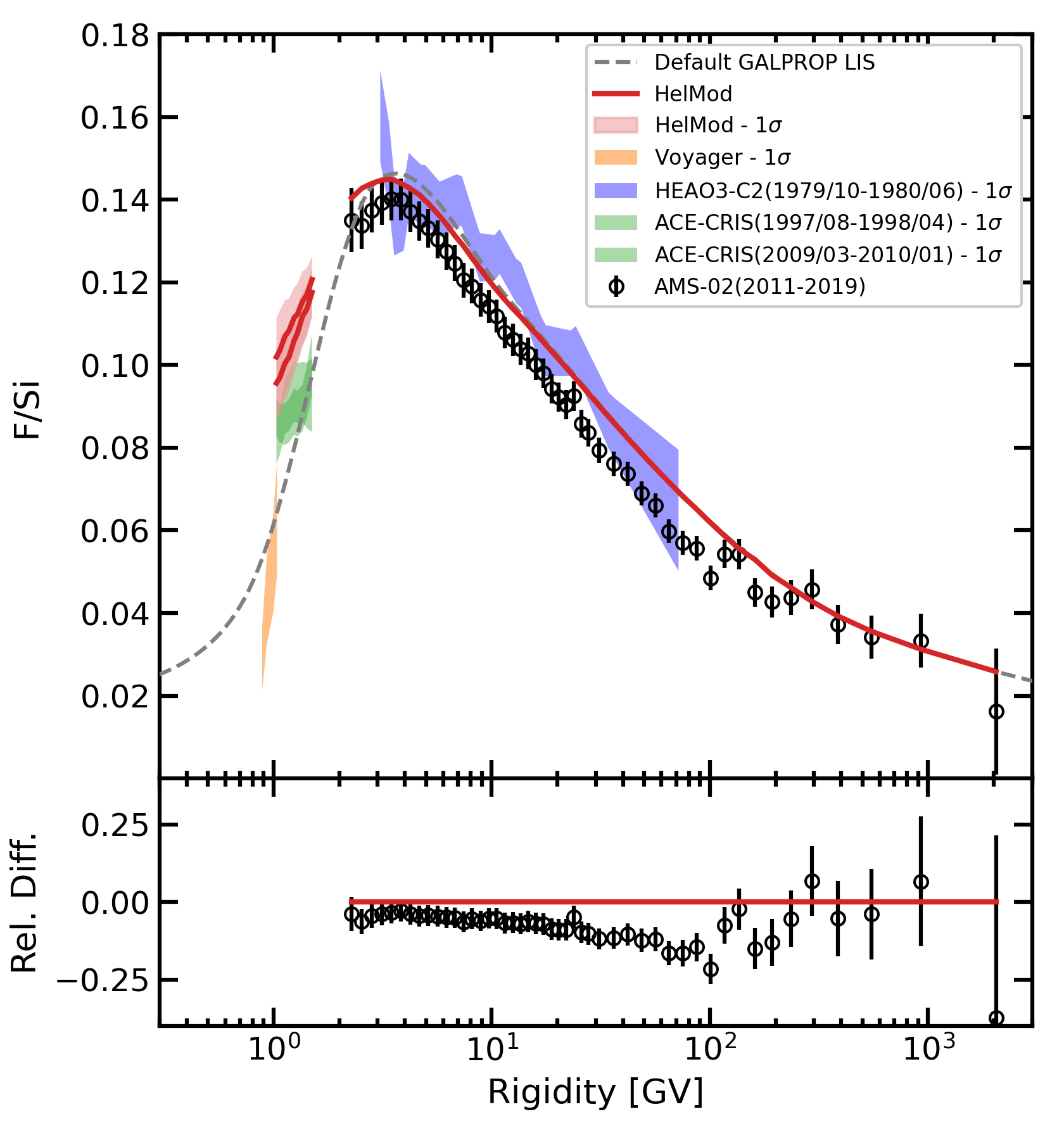}\hfill
	\includegraphics[width=0.49\textwidth]{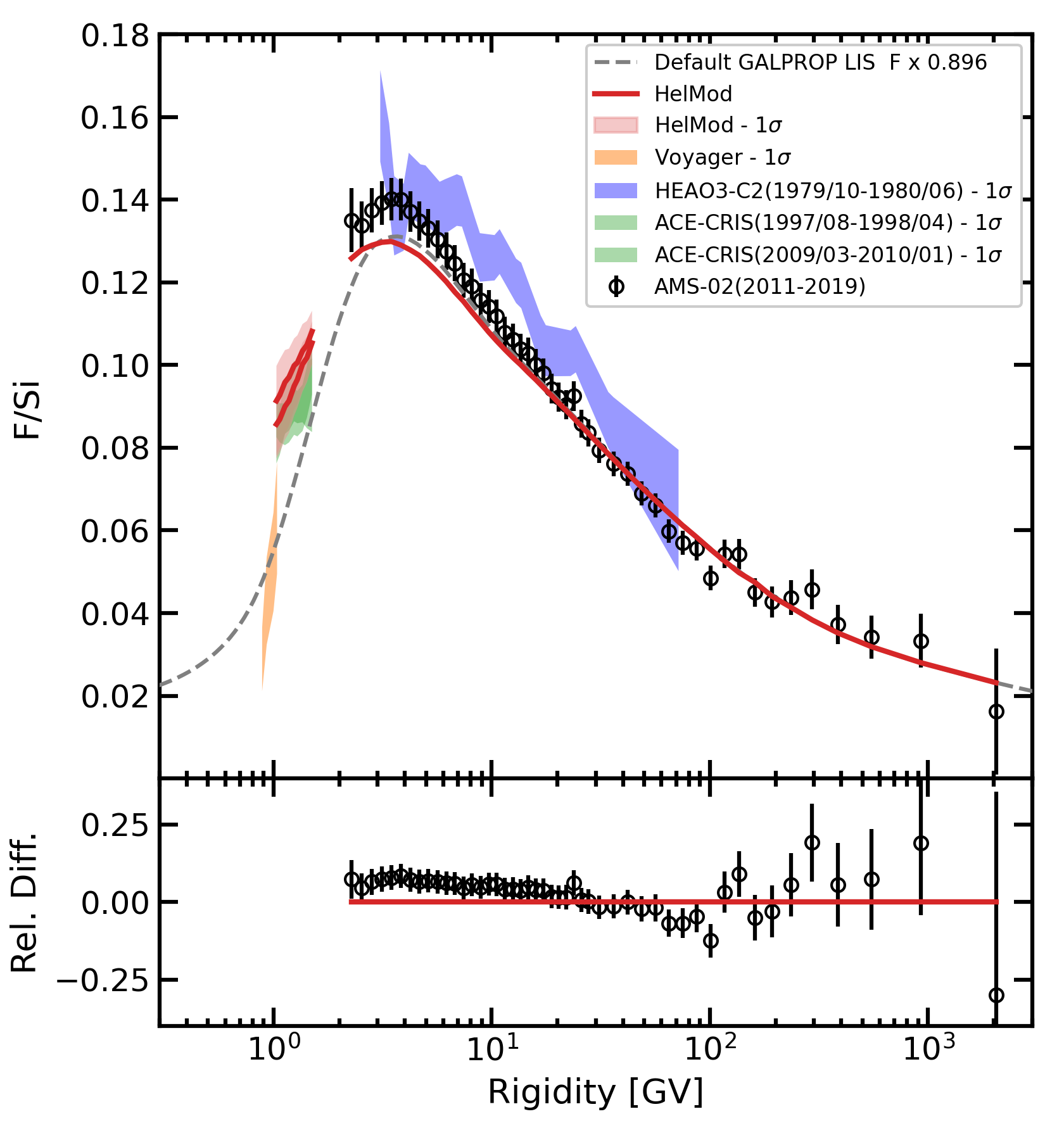}
	\includegraphics[width=0.49\textwidth]{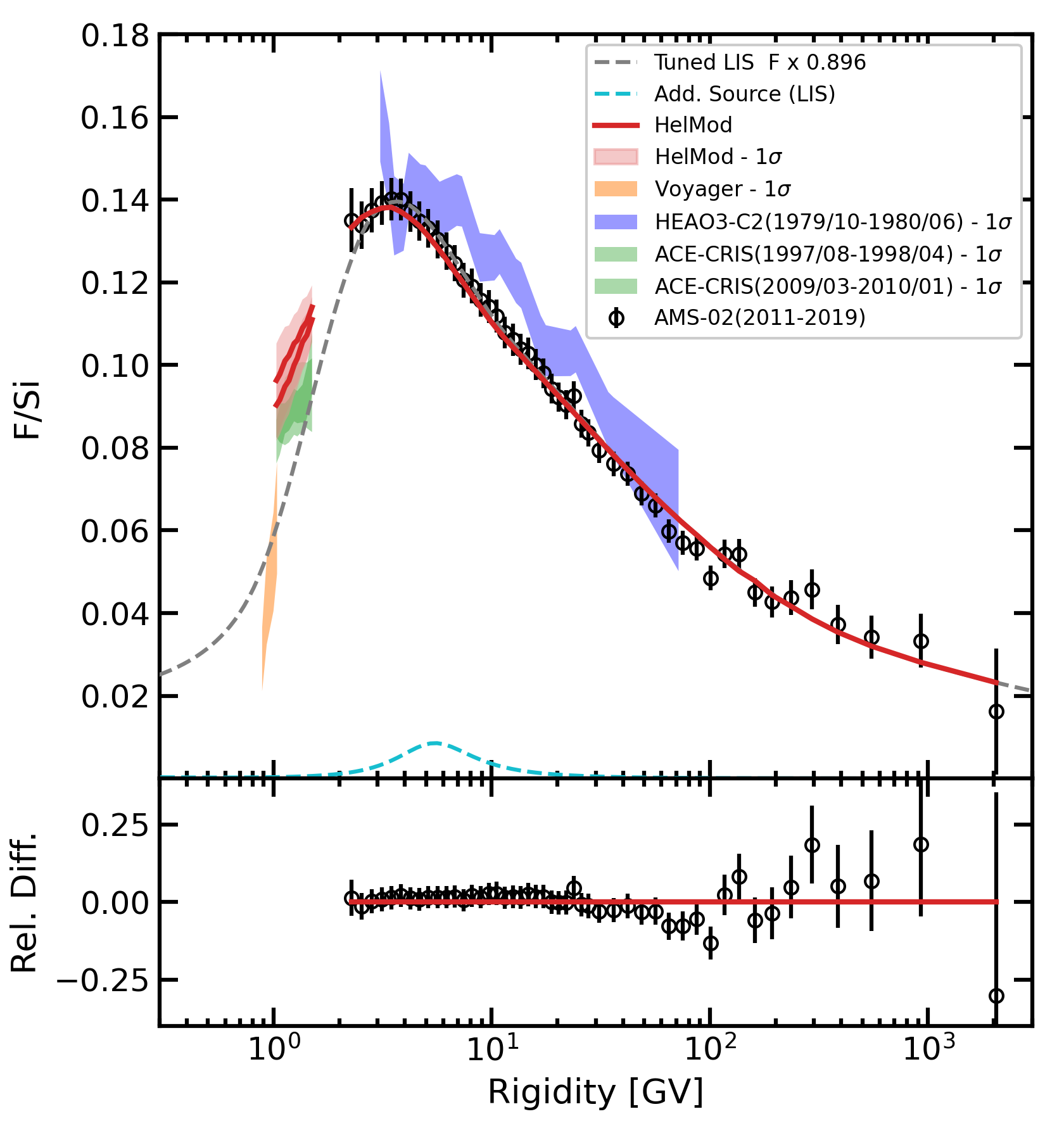}
	\caption{
{\it Top left:} The calculated default F/Si ratio as compared with Voyager~1 \citep{2016ApJ...831...18C}, ACE-CRIS \citep{2013ApJ...770..117L}, HEAO-3-C2 \citep{1990A&A...233...96E}, and AMS-02 data \citep{2020PhRvL.124u1102A, 2021PhRvL.126h1102A}.  {\it Top right:} The calculated default F/Si ratio is renormalized with a factor of 0.896.  {\it Bottom:} The same as in the top right, but with added primary fluorine component (see the injection spectrum in Table~\ref{tbl-inject}), where the total calculated fluorine spectrum is tuned to the data (Fig.~\ref{fig:F-spec}). The dashed cyan line shows separately the propagated ratio with primary fluorine only. In all panels, the dashed gray line shows the LIS ratio, and the solid red line is the corresponding modulated ratio. In all plots, the Voyager 1, ACE-CRIS, and HEAO-3-C2 data are converted from kinetic energy per nucleon to rigidity assuming $A/Z$=2 for Si. These data are shown as shaded areas with the width corresponding to 1$\sigma$ error.
Each panel also shows the relative difference between our calculations and the AMS-02 data set.
	}
	\label{fig:FSi-ratios}
\end{figure*}

\begin{figure*}[tp!]%[tb!]
	\centering
	\includegraphics[width=0.49\textwidth]{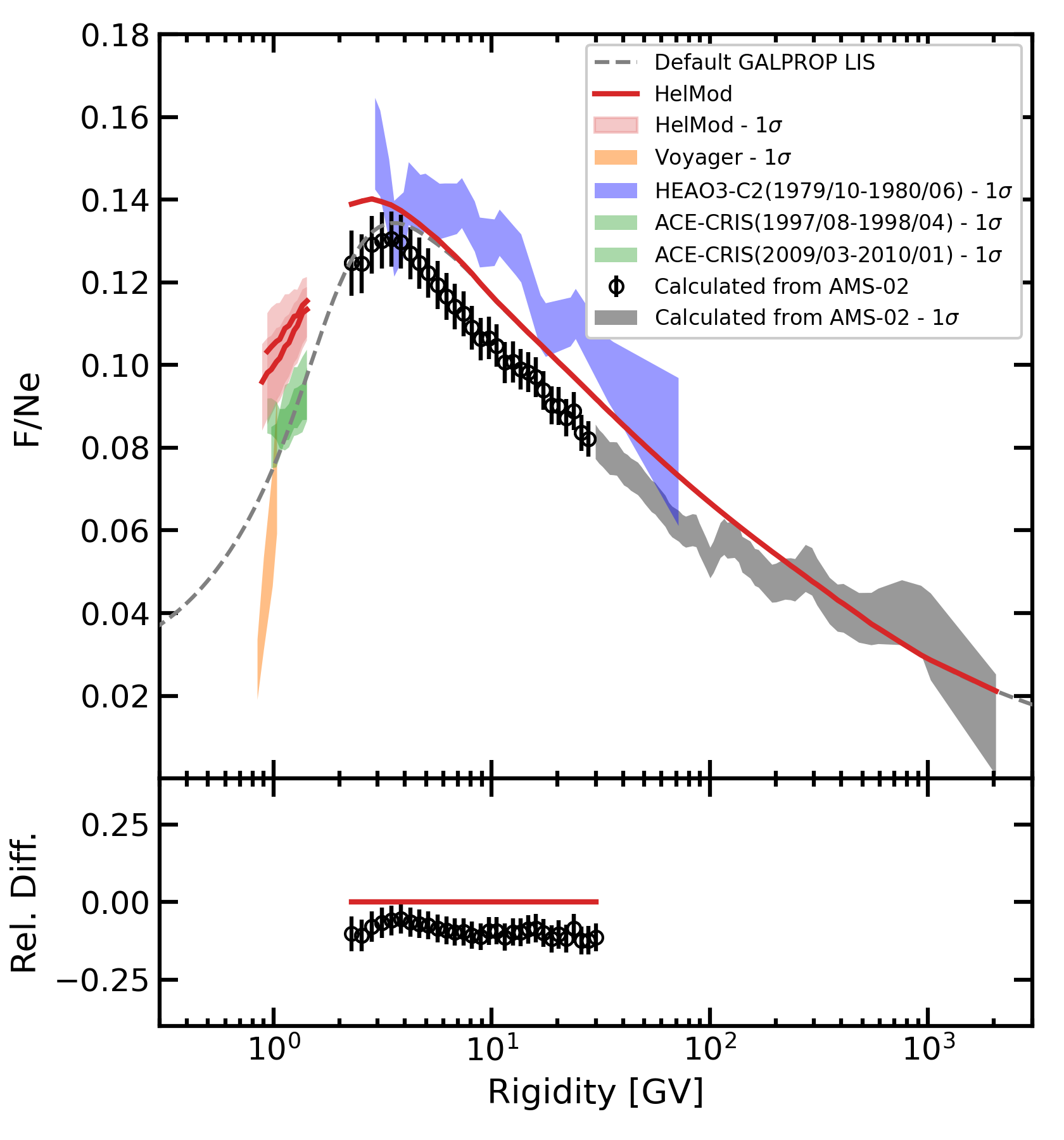}\hfill
	\includegraphics[width=0.49\textwidth]{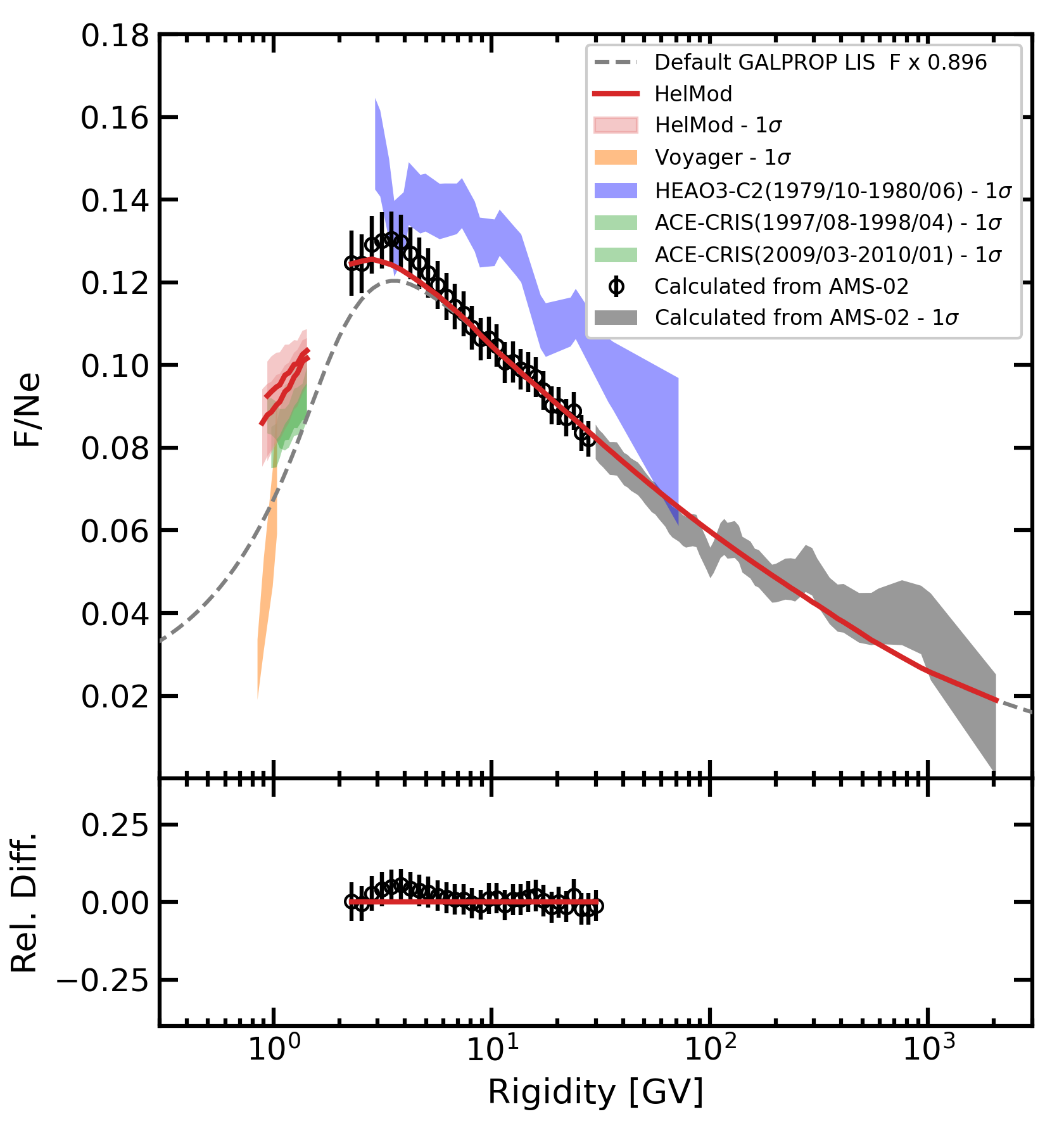}
	\includegraphics[width=0.49\textwidth]{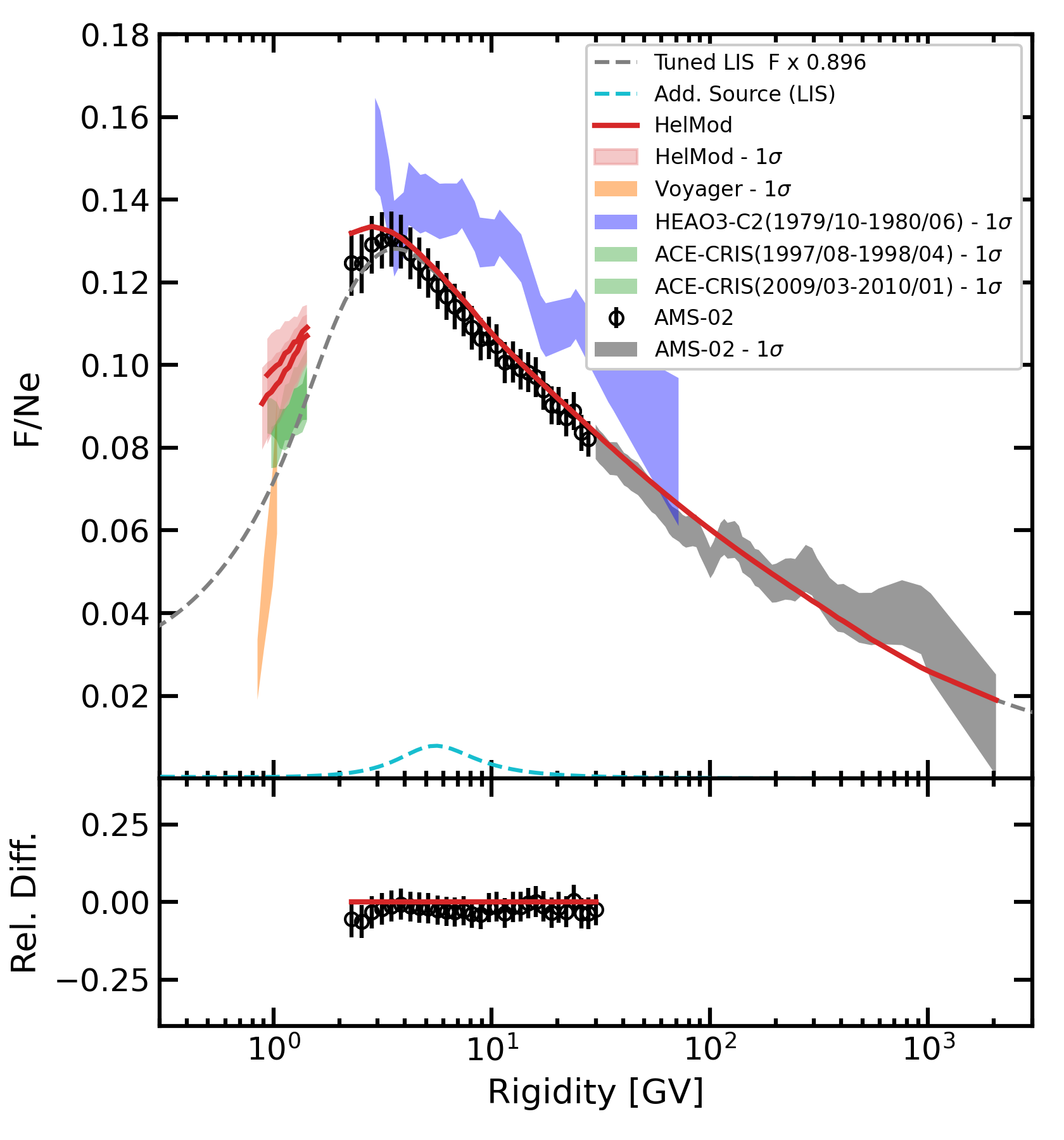}
	\caption{
{\it Top left:} The calculated default F/Ne ratio as compared with Voyager~1 \citep{2016ApJ...831...18C}, ACE-CRIS \citep{2013ApJ...770..117L}, HEAO-3-C2 \citep{1990A&A...233...96E}, and AMS-02 data \citep{2020PhRvL.124u1102A, 2021PhRvL.126h1102A}.  {\it Top right:} The calculated default F/Ne ratio is renormalized with a factor of 0.896.  {\it Bottom:} The same as in the top right, but with added primary fluorine component (see the injection spectrum in Table~\ref{tbl-inject}), where the total calculated fluorine spectrum is tuned to the data (Fig.~\ref{fig:F-spec}). The dashed cyan line shows separately the propagated ratio with primary fluorine only. In all panels, the dashed gray line shows the LIS ratio, and the solid red line is the corresponding modulated ratio. In all plots, the Voyager 1, ACE-CRIS, and HEAO-3-C2 data are converted from kinetic energy per nucleon to rigidity assuming $A/Z$=2 for Ne. AMS-02 data shown as the data points for the identical rigidity bins for F and Ne fluxes ($<$30 GV), and interpolated where the rigidity binning is different ($>$30 GV). 
These data are shown as shaded areas with the width corresponding to 1$\sigma$ error.
Each panel also shows the relative difference between our calculations and the AMS-02 data set, where the data points are available.
	}
	\label{fig:FNe-ratios}
\end{figure*}

The calculated spectra of Ne, Mg, and Si, which are the main contributors to the production of secondary F, and the B/C ratio were tuned to the high precision measurements by Voyager~1, ACE-CRIS, and AMS-02 \citep[see Figs.\ 3, 4 in][]{2020ApJS..250...27B}. This eliminates possible systematic errors in calculation of the spectrum of secondary F due to the poorly constrained propagation parameters, especially in the rigidity range where the excess is observed.
In Figs.~\ref{fig:FSi-ratios} and \ref{fig:FNe-ratios} the calculated F/Si and F/Ne ratios are compared with Voyager~1 \citep{2016ApJ...831...18C}, ACE-CRIS \citep{2013ApJ...770..117L}, HEAO-3-C2 \citep{1990A&A...233...96E}, and AMS-02 data \citep{2020PhRvL.124u1102A, 2021PhRvL.126h1102A}. The top left panels show the calculated default F/Si and F/Ne ratios with secondary fluorine only. The top right panels show the default ratios with renormalized fluorine ($\times$0.896). The renormalized plots exhibit the excess below 10 GV, while still agreeing with ACE-CRIS and Voyager 1 data. Above this rigidity the agreement with AMS-02 data is good. The bottom panels show the calculated F/Si and F/Ne ratios with the renormalized fluorine ($\times$0.896) and with the primary propagated fluorine component tuned to the data (Figs.~\ref{fig:F-spec}, bottom panel). The dashed cyan line shows separately the propagated ratios with primary fluorine only. In all plots, the Voyager 1, ACE-CRIS, and HEAO-3-C2 data are converted from kinetic energy per nucleon to rigidity assuming $A/Z$=2 for Si and Ne. In Fig.~\ref{fig:FNe-ratios}, AMS-02 data shown as the data points for the identical rigidity bins for F and Ne fluxes ($<$30 GV), and interpolated where the rigidity binning is different ($>$30 GV). The shaded areas show the measured ratios with the width corresponding to 1$\sigma$ error.

Apart from the overall normalization, possibly attributable to insufficiently constrained fluorine production cross sections (which however become fairly flat $\gtrsim$1 GeV/nucleon), we see three possible reasons for the observed low-energy ($<$10 GV) excess: (i) an underestimate of the total inelastic cross section of fluorine, which has a broad minimum at a few 100 MeV nucleon$^{-1}$ (see, e.g., Fig.~1 in \citealt{1996PhRvC..54.1329W}), (ii) the effective diffusion coefficient in the rigidity range 1-10 GV is somewhat smaller than that derived from the B/C ratio, and (iii) a primary fluorine component. We note that the fine tuning of the low-rigidity part in the Ne, Mg, and Si spectra does not help to remove the excess in fluorine entirely though could make it somewhat less significant. We, therefore, proceed with our original Ne, Mg, and Si spectra that are not fine-tuned to the fluorine data.

Underestimation of the total inelastic cross section of fluorine as a primary reason for the excess can be excluded by consideration of the accelerator data. \citet{Bobchenko:1979hp} provide a table of measured proton-nucleus cross sections for the proton momenta from 5 GeV/c to 9 GeV/c that correspond to the ambient $^{19}$F rigidity range 11--19 GV in the inverse kinematics, while the measurement accuracy is stated as 1-2\%. The average value of the total inelastic cross section in this interval is 350.9 mb; the deviations do not exceed $\pm$5 mb with just two out of total seventeen points being clear outliers at 361$\pm$5 mb and 358$\pm$5 mb. This is the energy range where the total inelastic cross sections are flat and they are easier to measure than the isotopic production. Besides, the total fluorine fragmentation cross section can be easily scaled from the nearby oxygen and neon cross sections. The parameterizations of the total inelastic cross sections used in the calculations \citep[e.g.,][]{BarPol1994} are tuned to the available data. 
Therefore, given a weak rigidity dependence of the diffusion coefficient, a possible error in the total inelastic cross section would be equivalent to a simple renormalization of the fluorine flux, very similar to the renormalization of the production cross sections.
That makes a significant error at around 5 GV, the maximum in the fluorine excess, rather unlikely. We note the absence of a similar excess in the spectra of neighboring nuclei, such as O, Ne, Mg, Si \citep{2020ApJS..250...27B}.

Variations in the diffusion coefficient seem unlikely as well. The ratio F/Si is the Si-group analog of the widely used B/C or B/O ratio for the light species. Both the F/Si and F/Ne ratios are consistent with predictions based on the propagation parameters derived from the B/C ratio above 10 GV. However, they demonstrate $\approx$6-7\% deviation in the rigidity range 3--7 GV. In terms of the diffusion coefficient, it would mean that the effective diffusion coefficient in this rigidity range probed by the F/Si and F/Ne ratios is about 10\% smaller than that in the case of the B/C ratio. Assuming the same distribution of sources of CR C, O, Ne, Mg, and Si nuclei, and given that the total inelastic cross sections of $^{16}$O (320 mb) and $^{20}$Ne (380 mb) are about the same, with neon being the main contributor to CR fluorine (see Fig.~1 in \citealt{2013ICRC...33..803M} and  Appendix C.1 in \citealt{2020ApJS..250...27B}), the B/C, F/Si, and F/Ne ratios are probing essentially the same Galactic volume. The total inelastic cross section of $^{28}$Si (465 mb) is only a factor of $\sim$1.3 larger, which corresponds to the effective propagation distance of a factor of $\sim$0.90 smaller than that of O and Ne \citep{2016ApJ...824...16J}. Here the values are given for the energy range of the excess, i.e.\ at a few GeV nucleon$^{-1}$. 

Meanwhile, we note that the measured spectra of the light primary nuclei He, C, O are flatter than the spectra of the Si-group nuclei, Ne, Mg, Si, with the difference in indices being $\sim$0.045 above 90 GV \citep{2020PhRvL.124u1102A}. If the difference in indices between O and Si nuclei is confirmed with larger statistics, it may indicate a somewhat different distribution of their sources. Besides that, measurements of other species, such as Na and Al spectra, P/S ratio, and especially the ratio of the iron group nuclei (Sc+V)/Fe, can be used to probe the variations of the diffusion coefficient in the local Galaxy.  

An interesting possibility would be a presence of the primary fluorine at low energies. We already discussed a hint for the primary fluorine in CRs in \citet[][Appendix C.1]{2020ApJS..250...27B}, but the large error bars and large scattering of the HEAO-3-C2 data points made this result inconclusive. The first ever accurate measurement of the fluorine spectrum in the rigidity range from 2 GV--2 TV \citep{2021PhRvL.126d1104A} allows the fluorine spectrum to be analyzed in finer details, and it seems that the excess also appears in the new data.

We do not discuss here a possibility of instrumental errors as they are rather unlikely. The AMS-02 detector, with its multiple redundancies, provides the state-of-the-art measurements of the CR species \citep{2021PhR...894....1A}. Before its launch, AMS-02 was tested extensively at the CERN test beam with electrons, positrons, protons, and pions. During more than 10 years of AMS-02 construction, a large international group of physicists have developed a comprehensive Monte Carlo simulation program for AMS-02. Importantly, the fragmentation cross sections of CR species inside the instrument are measured by the AMS-02 itself using the silicon tracker layers. The procedure and survival probabilities for different elements are described in \citet{2021PhR...894....1A}. The data analysis is performed in parallel by several groups, which employ different methods. A good agreement between their results is required before a paper is submitted for a publication.

In the Appendix, we provide an analytical parameterization of the secondary-only renormalized fluorine LIS, Eq.~(\ref{eq:F}), and the total fluorine LIS, Eq.~(\ref{eq:F_primes}), that includes also the primary component. The parameters are provided in Tables \ref{Tbl-1-10} and \ref{Tbl-1-10-bis}, correspondingly. We also provide numerical tables for both cases, which tabulate the LIS in rigidity $R$ (Table \ref{Tbl-FluorineLIS-Rigi}) and in kinetic energy $E_{\rm kin}$ per nucleon (Table \ref{Tbl-FluorineLIS-EKin}) for the secondary-only fluorine LIS, and corresponding Tables \ref{Tbl-FluorineLIS-Rigi_primes}, \ref{Tbl-FluorineLIS-EKin_primes} for the total fluorine LIS. The analytical parameterizations, Eqs.~(\ref{eq:F}), (\ref{eq:F_primes}) and numerical tables include a rigidity-independent normalization factor, $a$=0.896.

\section{Primary Fluorine}\label{fluorine}
%%%%%%%%%%%%%%%%%%%

Precise CR measurements over the last decade have lead to the discoveries of new features in spectra of CR species, various breaks and excesses in the energy range from sub-GeV to multi-TeV. Using combined measurements of AMS-02, Voyager 1, and ACE-CRIS we were able to establish a presence of primary lithium in CRs \citep{2020ApJ...889..167B}, and to discover the low-energy excess in iron \citep{2021ApJ...913....5B}. In the present paper, we argue that the fluorine spectrum may also exhibit a low-energy excess. These excesses may harbor the keys to understanding our local Galactic environment. Here we discuss potential sources of primary CR fluorine.  

The ISM abundance of fluorine is anomalously low because it is easily destroyed in stars through either $p$- or $\alpha$-captures. Its solar system abundance relative to oxygen ranges from (F/O)$_\odot = 0.74\times10^{-4}$ in the solar photosphere to $1.05\times10^{-4}$ in meteorites \citep{2009ARA&A..47..481A}. This can be compared with the derived F/O ratio in CR sources: (F/O)$_{\rm srs} = 4\times10^{-4}$ or [F/O]$_{\rm srs}=0.6-0.7$ dex dependently on the (F/O)$_\odot$ used, where [X/Y] = $\log_{10}$(X/Y)--$\log_{10}$(X/Y)$_\odot$. Here we integrated over the injection spectra of primary fluorine (Table~\ref{tbl-inject}) and oxygen \citep[Tables 2, 3 in][]{2020ApJS..250...27B} above 1 GV, where the excess is observed. If integrated above 0.1 GV, then the ratio becomes (F/O)$_{\rm srs}$ $= 1.7\times10^{-4}$ or [F/O]$_{\rm srs}=0.21-0.36$ dex, closer to the solar system abundance.

The origin of cosmic fluorine is still not well constrained. Several nucleosynthetic channels at different phases of stellar evolution have been suggested, but these must be constrained by observations. The main astrophysical sources of fluorine are thought to be supernovae Type II (SN II), Wolf--Rayet (WR) stars, and the asymptotic giant branch (AGB) of intermediate-mass stars \citep[e.g.,][]{2000A&A...355..176M, 2004MNRAS.354..575R, 2019MNRAS.490.4307O}. These sources become important at different stages of chemical evolution of the Galaxy, with the $\nu$-spallation of neon in SN II dominating at early times in low metallicity environment, while WR and AGB stars dominating at the later stages at solar and supersolar metallicities. The calculations show that reaching the solar level of [F/O] = 0 at the present epoch requires all three types of sources to contribute \citep{2004MNRAS.354..575R}. 

Meanwhile, determining the fluorine abundance in stars is challenging. A diagnostic is via observations of molecular HF emission lines in the K and N bands (2.1--2.4 $\mu$m and 8--13 $\mu$m, respectively), which are observable only in cool giants ($T_{\rm eff}<4500$ K). Analysis of the spectra of K giants show that the fluorine-to-oxygen abundance ratio [F/O] increases as a function of oxygen abundance [O/H] with a slope 1$\pm$0.2, and can reach [F/O]$\approx$0.4 dex for [O/H]$\approx$0.2 dex \citep{2020ApJ...893...37R}. Behavior like this is typical for the so-called ``secondary'' elements, whose synthesis in stars involves nuclides pre-existing in the ISM gas of which stars form. A corresponding abundance ratio for the so-called ``primary'' elements would be a constant, because the abundances of primary elements increase proportionally to metallicity. An example of a primary element is oxygen.

The key isotope that contributes to the production of fluorine in different environments is $^{14}$N, which is converted to fluorine via a series of proton and $\alpha$ captures $^{14}$N($\alpha$, $\gamma$)$^{18}$F ($\beta^+$)$^{18}$O($p$,$\alpha$)$^{15}$N($\alpha$,$\gamma$)$^{19}$F in the presence of $^{13}$C; the latter is needed for the generation of protons \citep{1990nuas.symp...60G, 2018MNRAS.476.3432P}. Such reactions can take place in rapidly rotating massive stars, thermally pulsating AGB stars, and massive WR stars. Novae are also proposed as a possible source of fluorine through proton captures by $^{17}$O. A useful review of fluorine nucleosynthesis is given in \citet{2020ApJ...893...37R}. 

A possible identification of a primary CR fluorine component would provide another (indirect) way of studying the synthesis of this species in CR sources. If true, the fluorine overabundance in the CR sources ([F/O]$_{\rm srs}=0.6-0.7$ dex above 1 GV or $0.21-0.36$ dex above 0.1 GV) could hint at perhaps yet unknown processes acting at certain stages of stellar evolution. Important clues to the origin of primary fluorine, and the nature of CR sources in the local Galaxy, can be deduced from observation (or non-observation) of low-energy excesses in the spectra of other mostly secondary elements of the Si-group, such as Na and Al.

\section{Conclusion} \label{conclusion}
%%%%%%%%%%%%%%%%%%%%%%%%%%%%%%%%%%%%%%%%%%%%%%%%%%%
%%%%%%%%%%%%%%%%%%%%%%%%%%%%%%%%%%%%%%%%%%%%%%%%%%%

Using the combined data of AMS-02 \citep{2021PhRvL.126d1104A}, ACE-CRIS \citep{2013ApJ...770..117L}, and Voyager 1 \citep{2016ApJ...831...18C} we analyzed the spectrum of fluorine over a wide rigidity range from MV to 2 TV. We found moderate discrepancies with the predictions made with the \galprop{}-\helmod{} framework. 

First, the default normalization of the whole spectrum was found to be off by $\sim$10\%. We attribute this discrepancy to the errors in the isotopic production cross sections that is well inside of the their typical uncertainties. Second, the renormalized fluorine spectrum and renormalized F/Si and F/Ne ratios show an excess below 10 GV. We discuss possible origin of this excess and rule out the underestimate of the total inelastic cross section of fluorine and the variations in the diffusion coefficient as the primary reasons for the excess. 

%%%%%%%%%%%%
We conclude that a contribution of primary fluorine could not be ruled out at this point, but its confirmation requires more data. We present the derived injection spectrum of the primary fluorine and discuss its main astrophysical sources. As in all our previous papers, we also provide an analytical parameterization and numerical tables of the fluorine LIS.

\acknowledgements
We are grateful to the anonymous referee for their thorough reading of the manuscript and useful remarks. Special thanks to Pavol Bobik, Giuliano Boella, Karel Kudela, Marian Putis, and Mario Zannoni for their continuous support of the \helmod{} project and many useful suggestions. This work is supported by ASI (Agenzia Spaziale Italiana) through a contract ASI-INFN No.\ 2019-19-HH.0 and by ESA (European Space Agency) through a contract 4000116146/16/NL/HK. Igor V.\ Moskalenko and Troy A.\ Porter acknowledge support from NASA Grant No.~NNX17AB48G. This research has made use of the SSDC Cosmic rays database \citep{2017ICRC...35.1073D} and LPSC Database of Charged Cosmic Rays \citep{2014A&A...569A..32M}. 

\bibliography{bibliography}

\appendix

\section{Analytical parameterization and numerical tables of the fluorine LIS}

Here we provide an analytical parameterization of the secondary-only renormalized fluorine LIS: 
\begin{align}
	% -------------------------------------------------- Z = 6 Fluorine ---------------------------------------------
	\label{eq:F}
	F_{\rm sec} (R)  = a \times %\\
	&\begin{cases}
		\displaystyle  b + cR + dR^5 - fR^2 - gG(R)\sqrt{R} - hR^2 G(R) - iRG^2(R),  & R\le 2.5\, {\rm GV}, \smallskip\\
		\displaystyle  R^{-2.7} \left[ -\frac{l}{R} + mR^2 + \frac{n}{o + R} + \frac{p}{q + rR} + \frac{s}{R \tanh\left[{o}/{(t + R)}\right]} - u - vR \right], & R> 2.5\, {\rm GV},
	\end{cases} 
\end{align}
where $R$ is the particle rigidity in GV, the values of the fitting parameters from $a$ to $z$ are given in Table \ref{Tbl-1-10}, and the function $G(x)$ is defined as: 
\[
 G(x)=e^{-x^2}
\]
The analytical representation, Eq.~(\ref{eq:F}), is also complemented by numerical tables calculated for the {\it I}-scenario, which tabulate the LIS in rigidity $R$ (Table \ref{Tbl-FluorineLIS-Rigi}) and in kinetic energy $E_{\rm kin}$ per nucleon (Table \ref{Tbl-FluorineLIS-EKin}).

\renewcommand\floatpagefraction{1.0}
\enlargethispage{50\baselineskip}

We also provide an analytical parameterization of the total fluorine LIS, which includes the primary component: 
\begin{align}
	% -------------------------------------------------- Z = 6 Fluorine with Primaries ---------------------------------------------
	\label{eq:F_primes}
	F_{\rm tot} (R)  = a\times 
	&\begin{cases}
		\displaystyle  o - pR  - qR\log{R} - rG(R)  + sG^3(R) + tR^2G(R) - uG[v + zG(R)], & R\le 2.5\, {\rm GV}, \smallskip\\
        \displaystyle  R^{-2.7} \left[ b+ cR + dR^2 - f\sqrt{R} - gR\sqrt{R} -\frac{h}{R} -\frac{i}{\sqrt{R}} + \frac{l}{m + R} + \frac{n}{R + \sqrt{R}}   \right],    &R> 2.5\, {\rm GV},    
\end{cases}
\end{align}
where $R$ is the particle rigidity in GV, the values of the fitting parameters from $a$ to $z$ are given in Table \ref{Tbl-1-10-bis}, and the function $G(x)$ is defined above.  The analytical representation, Eq.~(\ref{eq:F_primes}), is also complemented by numerical tables calculated for the {\it I}-scenario, which tabulate the LIS in rigidity $R$ (Table \ref{Tbl-FluorineLIS-Rigi_primes}) and in kinetic energy $E_{\rm kin}$ per nucleon (Table \ref{Tbl-FluorineLIS-EKin_primes}).

\begin{deluxetable}{rlrlrlrl}[ht!]
 	\tablecolumns{8}
 	\tablewidth{0mm}
 	\tablecaption{Parameters of the fit to the secondary-only renormalized fluorine LIS, Eq.~(\ref{eq:F}) \label{Tbl-1-10}}
 	\tablehead{
 \colhead{Param}  &  \colhead{Value}  & \colhead{Param}  & \colhead{Value}  & \colhead{Param}  &  \colhead{Value}  & \colhead{Param}    &  \colhead{Value}
 	}
	\startdata
	$a$\phantom{a}   & 0.896e+0  & $g$\phantom{a}  & 2.2707e-2 & $n$\phantom{a} & 1.1615e+2 & $s$\phantom{a} & 6.2037e+1 \\
	$b$\phantom{a}   & 3.0304e-3 & $h$\phantom{a}  & 2.6461e-2 & $o$\phantom{a} & 2.1291e+1 & $t$\phantom{a} & 5.2771e+0 \\
	$c$\phantom{a}   & 1.4857e-1 & $i$\phantom{a}  & 9.9310e-2 & $p$\phantom{a} & 2.7873e+1 & $u$\phantom{a} & 2.5986e+0 \\
	$d$\phantom{a}   & 4.7068e-4 & $l$\phantom{a}  & 6.6451e+1 & $q$\phantom{a} & 1.7222e+2 & $v$\phantom{a} & 2.0513e-7 \\
	$f$\phantom{a}   & 5.5235e-2 & $m$\phantom{a}  & 2.0120e-13 & $r$\phantom{a} & 1.0152e-2 & \nodata & \nodata \\
	\enddata
\end{deluxetable}

\begin{deluxetable}{rlrlrlrl}[ht!]
\tablecolumns{8}
\tablewidth{0mm}
\tablecaption{Parameters of the fit to the total fluorine LIS, Eq.~(\ref{eq:F_primes}) \label{Tbl-1-10-bis}}
\tablehead{
	\colhead{Param}  &  \colhead{Value}  & \colhead{Param}  & \colhead{Value}  & \colhead{Param}  &  \colhead{Value}  & \colhead{Param}    &  \colhead{Value}
}
	\startdata
	$a$\phantom{a}   & 0.896e+0    & $g$\phantom{a}  & 1.16195e-8 & $n$\phantom{a} & 2.57541e+1 & $s$\phantom{a} & 4.77105e-2  \\
	$b$\phantom{a}   & 6.12968e-1  & $h$\phantom{a}  & 1.37360e+1 & $o$\phantom{a} & 1.81461e-1 & $t$\phantom{a} & 9.94025e-3  \\
	$c$\phantom{a}   & 7.19610e-6  & $i$\phantom{a}  & 4.02914e+0 & $p$\phantom{a} & 2.45030e-2 & $u$\phantom{a} & 2.03517e-2   \\
	$d$\phantom{a}   & 7.17705e-12 & $l$\phantom{a}  & 1.01968e+2 & $q$\phantom{a} & 9.31457e-3 & $v$\phantom{a} & 1.58760e-1  \\
	$f$\phantom{a}   & 2.22764e-3  & $m$\phantom{a}  & 4.69709e+1 & $r$\phantom{a} & 2.29360e-1 & $z$\phantom{a} & 1.98665e+0 \\
	\enddata
\end{deluxetable}

\begin{deluxetable}{cccccccccc}[tp!]
\tabletypesize{\footnotesize}
    \tablecolumns{10}
    \tablewidth{0mm}
    \tablecaption{$Z=9$ -- secondary-only renormalized fluorine LIS\label{Tbl-FluorineLIS-Rigi}}
    \tablehead{
        \colhead{Rigidity} & \colhead{Differential} &
        \colhead{Rigidity} & \colhead{Differential} &
        \colhead{Rigidity} & \colhead{Differential} &
        \colhead{Rigidity} & \colhead{Differential} &
        \colhead{Rigidity} & \colhead{Differential} \\ [-2ex] 
        \colhead{GV} & \colhead{intensity} &
        \colhead{GV} & \colhead{intensity} &
        \colhead{GV} & \colhead{intensity} &
        \colhead{GV} & \colhead{intensity} &
        \colhead{GV} & \colhead{intensity}  
    }
	\startdata
	9.114e-02 & 1.554e-04 & 7.359e-01 & 3.419e-02 & 1.018e+01 & 2.802e-03 & 5.322e+02 & 2.265e-08 & 3.346e+04 & 1.991e-13\\
	9.565e-02 & 1.781e-04 & 7.735e-01 & 3.778e-02 & 1.105e+01 & 2.249e-03 & 5.858e+02 & 1.720e-08 & 3.684e+04 & 1.517e-13\\
	1.004e-01 & 2.023e-04 & 8.132e-01 & 4.157e-02 & 1.200e+01 & 1.798e-03 & 6.449e+02 & 1.307e-08 & 4.057e+04 & 1.156e-13\\
	1.053e-01 & 2.299e-04 & 8.551e-01 & 4.554e-02 & 1.304e+01 & 1.432e-03 & 7.100e+02 & 9.948e-09 & 4.467e+04 & 8.814e-14\\
	1.105e-01 & 2.612e-04 & 8.992e-01 & 4.966e-02 & 1.419e+01 & 1.135e-03 & 7.816e+02 & 7.577e-09 & 4.919e+04 & 6.717e-14\\
	1.160e-01 & 2.969e-04 & 9.459e-01 & 5.386e-02 & 1.545e+01 & 8.963e-04 & 8.605e+02 & 5.775e-09 & 5.417e+04 & 5.119e-14\\
	1.217e-01 & 3.376e-04 & 9.952e-01 & 5.809e-02 & 1.684e+01 & 7.045e-04 & 9.474e+02 & 4.403e-09 & 5.965e+04 & 3.901e-14\\
	1.277e-01 & 3.839e-04 & 1.047e+00 & 6.231e-02 & 1.836e+01 & 5.512e-04 & 1.043e+03 & 3.359e-09 & 6.569e+04 & 2.973e-14\\
	1.341e-01 & 4.367e-04 & 1.103e+00 & 6.645e-02 & 2.004e+01 & 4.295e-04 & 1.148e+03 & 2.564e-09 & 7.233e+04 & 2.265e-14\\
	1.407e-01 & 4.967e-04 & 1.161e+00 & 7.042e-02 & 2.189e+01 & 3.333e-04 & 1.264e+03 & 1.958e-09 & 7.965e+04 & 1.726e-14\\
	1.476e-01 & 5.651e-04 & 1.223e+00 & 7.414e-02 & 2.392e+01 & 2.576e-04 & 1.392e+03 & 1.495e-09 & 8.771e+04 & 1.315e-14\\
	1.549e-01 & 6.431e-04 & 1.289e+00 & 7.753e-02 & 2.616e+01 & 1.983e-04 & 1.533e+03 & 1.142e-09 & 9.658e+04 & 1.002e-14\\
	1.626e-01 & 7.319e-04 & 1.358e+00 & 8.049e-02 & 2.862e+01 & 1.522e-04 & 1.688e+03 & 8.720e-10 & 1.064e+05 & 7.637e-15\\
	1.707e-01 & 8.330e-04 & 1.432e+00 & 8.292e-02 & 3.133e+01 & 1.164e-04 & 1.858e+03 & 6.659e-10 & 1.171e+05 & 5.819e-15\\
	1.791e-01 & 9.481e-04 & 1.511e+00 & 8.474e-02 & 3.431e+01 & 8.885e-05 & 2.046e+03 & 5.085e-10 & 1.290e+05 & 4.434e-15\\
	1.880e-01 & 1.079e-03 & 1.595e+00 & 8.588e-02 & 3.759e+01 & 6.764e-05 & 2.253e+03 & 3.884e-10 & 1.420e+05 & 3.379e-15\\
	1.973e-01 & 1.229e-03 & 1.684e+00 & 8.627e-02 & 4.121e+01 & 5.137e-05 & 2.481e+03 & 2.967e-10 & 1.564e+05 & 2.574e-15\\
	2.070e-01 & 1.399e-03 & 1.780e+00 & 8.581e-02 & 4.518e+01 & 3.894e-05 & 2.731e+03 & 2.267e-10 & 1.722e+05 & 1.962e-15\\
	2.173e-01 & 1.592e-03 & 1.881e+00 & 8.449e-02 & 4.957e+01 & 2.945e-05 & 3.008e+03 & 1.731e-10 & 1.896e+05 & 1.495e-15\\
	2.280e-01 & 1.813e-03 & 1.990e+00 & 8.232e-02 & 5.439e+01 & 2.224e-05 & 3.312e+03 & 1.322e-10 & 2.088e+05 & 1.139e-15\\
	2.393e-01 & 2.063e-03 & 2.107e+00 & 7.934e-02 & 5.970e+01 & 1.677e-05 & 3.647e+03 & 1.010e-10 &  \nodata & \nodata \\
	2.512e-01 & 2.349e-03 & 2.232e+00 & 7.563e-02 & 6.555e+01 & 1.262e-05 & 4.015e+03 & 7.709e-11 &  \nodata & \nodata \\
	2.637e-01 & 2.673e-03 & 2.366e+00 & 7.139e-02 & 7.199e+01 & 9.489e-06 & 4.421e+03 & 5.886e-11 &  \nodata & \nodata \\
	2.767e-01 & 3.041e-03 & 2.511e+00 & 6.670e-02 & 7.908e+01 & 7.126e-06 & 4.869e+03 & 4.493e-11 &  \nodata & \nodata \\
	2.905e-01 & 3.460e-03 & 2.666e+00 & 6.164e-02 & 8.688e+01 & 5.346e-06 & 5.361e+03 & 3.429e-11 &  \nodata & \nodata \\
	3.049e-01 & 3.934e-03 & 2.833e+00 & 5.632e-02 & 9.548e+01 & 4.007e-06 & 5.903e+03 & 2.617e-11 &  \nodata & \nodata \\
	3.201e-01 & 4.473e-03 & 3.014e+00 & 5.090e-02 & 1.049e+02 & 3.001e-06 & 6.501e+03 & 1.997e-11 &  \nodata & \nodata \\
	3.360e-01 & 5.083e-03 & 3.210e+00 & 4.555e-02 & 1.154e+02 & 2.245e-06 & 7.158e+03 & 1.524e-11 &  \nodata & \nodata \\
	3.527e-01 & 5.774e-03 & 3.421e+00 & 4.037e-02 & 1.268e+02 & 1.678e-06 & 7.882e+03 & 1.163e-11 &  \nodata & \nodata \\
	3.702e-01 & 6.556e-03 & 3.650e+00 & 3.545e-02 & 1.395e+02 & 1.254e-06 & 8.679e+03 & 8.870e-12 &  \nodata & \nodata \\
	3.887e-01 & 7.439e-03 & 3.899e+00 & 3.085e-02 & 1.534e+02 & 9.365e-07 & 9.558e+03 & 6.766e-12 &  \nodata & \nodata \\
	4.081e-01 & 8.435e-03 & 4.169e+00 & 2.660e-02 & 1.687e+02 & 6.995e-07 & 1.052e+04 & 5.161e-12 &  \nodata & \nodata \\
	4.285e-01 & 9.559e-03 & 4.463e+00 & 2.275e-02 & 1.856e+02 & 5.225e-07 & 1.159e+04 & 3.936e-12 &  \nodata & \nodata \\
	4.499e-01 & 1.082e-02 & 4.783e+00 & 1.932e-02 & 2.042e+02 & 3.904e-07 & 1.276e+04 & 3.002e-12 &  \nodata & \nodata \\
	4.724e-01 & 1.224e-02 & 5.132e+00 & 1.633e-02 & 2.247e+02 & 2.919e-07 & 1.405e+04 & 2.289e-12 &  \nodata & \nodata \\
	4.961e-01 & 1.383e-02 & 5.513e+00 & 1.371e-02 & 2.472e+02 & 2.184e-07 & 1.547e+04 & 1.746e-12 &  \nodata & \nodata \\
	5.210e-01 & 1.560e-02 & 5.929e+00 & 1.145e-02 & 2.720e+02 & 1.636e-07 & 1.704e+04 & 1.331e-12 &  \nodata & \nodata \\
	5.472e-01 & 1.758e-02 & 6.384e+00 & 9.519e-03 & 2.993e+02 & 1.227e-07 & 1.876e+04 & 1.015e-12 &  \nodata & \nodata \\
	\enddata
    \tablecomments{Differential Intensity units: (m$^2$ s sr GV)$^{-1}$.}
\end{deluxetable}

\begin{deluxetable}{cccccccccc}%[p!]
\tabletypesize{\footnotesize}
    \tablecolumns{10}
    \tablewidth{0mm}
    \tablecaption{$Z=9$ -- secondary-only renormalized fluorine LIS\label{Tbl-FluorineLIS-EKin}}
    \tablehead{
        \colhead{$E_{\rm kin}$} & \colhead{Differential} &
        \colhead{$E_{\rm kin}$} & \colhead{Differential} &
        \colhead{$E_{\rm kin}$} & \colhead{Differential} &
        \colhead{$E_{\rm kin}$} & \colhead{Differential} &
        \colhead{$E_{\rm kin}$} & \colhead{Differential} \\ [-2ex] 
        \colhead{GeV/n} & \colhead{intensity} &
        \colhead{GeV/n} & \colhead{intensity} &
        \colhead{GeV/n} & \colhead{intensity} &
        \colhead{GeV/n} & \colhead{intensity} &
        \colhead{GeV/n} & \colhead{intensity}  
    }
	\startdata
1.000e-03 & 7.085e-03 & 6.309e-02 & 2.059e-01 & 3.981e+00 & 6.024e-03 & 2.512e+02 & 4.782e-08 & 1.585e+04 & 4.203e-13\\
1.101e-03 & 7.739e-03 & 6.948e-02 & 2.179e-01 & 4.384e+00 & 4.822e-03 & 2.766e+02 & 3.631e-08 & 1.745e+04 & 3.203e-13\\
1.213e-03 & 8.378e-03 & 7.651e-02 & 2.296e-01 & 4.827e+00 & 3.847e-03 & 3.046e+02 & 2.760e-08 & 1.922e+04 & 2.441e-13\\
1.335e-03 & 9.073e-03 & 8.425e-02 & 2.411e-01 & 5.315e+00 & 3.057e-03 & 3.354e+02 & 2.100e-08 & 2.116e+04 & 1.861e-13\\
1.470e-03 & 9.826e-03 & 9.277e-02 & 2.521e-01 & 5.853e+00 & 2.420e-03 & 3.693e+02 & 1.600e-08 & 2.330e+04 & 1.418e-13\\
1.619e-03 & 1.065e-02 & 1.022e-01 & 2.623e-01 & 6.446e+00 & 1.907e-03 & 4.067e+02 & 1.219e-08 & 2.566e+04 & 1.081e-13\\
1.783e-03 & 1.154e-02 & 1.125e-01 & 2.716e-01 & 7.098e+00 & 1.497e-03 & 4.478e+02 & 9.296e-09 & 2.825e+04 & 8.235e-14\\
1.963e-03 & 1.250e-02 & 1.239e-01 & 2.798e-01 & 7.816e+00 & 1.170e-03 & 4.931e+02 & 7.091e-09 & 3.111e+04 & 6.275e-14\\
2.162e-03 & 1.355e-02 & 1.364e-01 & 2.868e-01 & 8.607e+00 & 9.110e-04 & 5.430e+02 & 5.413e-09 & 3.426e+04 & 4.782e-14\\
2.381e-03 & 1.469e-02 & 1.502e-01 & 2.924e-01 & 9.478e+00 & 7.065e-04 & 5.980e+02 & 4.133e-09 & 3.773e+04 & 3.644e-14\\
2.622e-03 & 1.594e-02 & 1.654e-01 & 2.964e-01 & 1.044e+01 & 5.457e-04 & 6.585e+02 & 3.156e-09 & 4.155e+04 & 2.777e-14\\
2.887e-03 & 1.728e-02 & 1.822e-01 & 2.986e-01 & 1.149e+01 & 4.199e-04 & 7.251e+02 & 2.411e-09 & 4.575e+04 & 2.116e-14\\
3.179e-03 & 1.875e-02 & 2.006e-01 & 2.990e-01 & 1.266e+01 & 3.221e-04 & 7.985e+02 & 1.841e-09 & 5.038e+04 & 1.612e-14\\
3.501e-03 & 2.034e-02 & 2.209e-01 & 2.973e-01 & 1.394e+01 & 2.463e-04 & 8.793e+02 & 1.406e-09 & 5.548e+04 & 1.228e-14\\
3.855e-03 & 2.207e-02 & 2.432e-01 & 2.936e-01 & 1.535e+01 & 1.879e-04 & 9.682e+02 & 1.073e-09 & 6.109e+04 & 9.361e-15\\
4.245e-03 & 2.395e-02 & 2.678e-01 & 2.878e-01 & 1.690e+01 & 1.430e-04 & 1.066e+03 & 8.200e-10 & 6.727e+04 & 7.133e-15\\
4.675e-03 & 2.599e-02 & 2.949e-01 & 2.800e-01 & 1.861e+01 & 1.086e-04 & 1.174e+03 & 6.264e-10 & 7.408e+04 & 5.435e-15\\
5.148e-03 & 2.820e-02 & 3.248e-01 & 2.700e-01 & 2.049e+01 & 8.228e-05 & 1.293e+03 & 4.785e-10 & 8.157e+04 & 4.141e-15\\
5.669e-03 & 3.061e-02 & 3.577e-01 & 2.580e-01 & 2.257e+01 & 6.223e-05 & 1.424e+03 & 3.655e-10 & 8.983e+04 & 3.155e-15\\
6.242e-03 & 3.322e-02 & 3.938e-01 & 2.443e-01 & 2.485e+01 & 4.698e-05 & 1.568e+03 & 2.791e-10 & 9.892e+04 & 2.404e-15\\
6.874e-03 & 3.606e-02 & 4.337e-01 & 2.291e-01 & 2.736e+01 & 3.541e-05 & 1.726e+03 & 2.132e-10 &  \nodata & \nodata \\
7.569e-03 & 3.913e-02 & 4.776e-01 & 2.128e-01 & 3.013e+01 & 2.666e-05 & 1.901e+03 & 1.628e-10 &  \nodata & \nodata \\
8.335e-03 & 4.246e-02 & 5.259e-01 & 1.960e-01 & 3.318e+01 & 2.004e-05 & 2.093e+03 & 1.243e-10 &  \nodata & \nodata \\
9.179e-03 & 4.607e-02 & 5.791e-01 & 1.789e-01 & 3.654e+01 & 1.505e-05 & 2.305e+03 & 9.485e-11 &  \nodata & \nodata \\
1.011e-02 & 4.998e-02 & 6.377e-01 & 1.617e-01 & 4.023e+01 & 1.129e-05 & 2.539e+03 & 7.240e-11 &  \nodata & \nodata \\
1.113e-02 & 5.421e-02 & 7.022e-01 & 1.447e-01 & 4.431e+01 & 8.461e-06 & 2.795e+03 & 5.525e-11 &  \nodata & \nodata \\
1.226e-02 & 5.878e-02 & 7.733e-01 & 1.283e-01 & 4.879e+01 & 6.336e-06 & 3.078e+03 & 4.216e-11 &  \nodata & \nodata \\
1.350e-02 & 6.372e-02 & 8.515e-01 & 1.128e-01 & 5.373e+01 & 4.739e-06 & 3.390e+03 & 3.217e-11 &  \nodata & \nodata \\
1.486e-02 & 6.905e-02 & 9.377e-01 & 9.830e-02 & 5.916e+01 & 3.543e-06 & 3.733e+03 & 2.455e-11 &  \nodata & \nodata \\
1.637e-02 & 7.480e-02 & 1.033e+00 & 8.501e-02 & 6.515e+01 & 2.647e-06 & 4.110e+03 & 1.873e-11 &  \nodata & \nodata \\
1.802e-02 & 8.099e-02 & 1.137e+00 & 7.293e-02 & 7.174e+01 & 1.977e-06 & 4.526e+03 & 1.428e-11 &  \nodata & \nodata \\
1.985e-02 & 8.764e-02 & 1.252e+00 & 6.209e-02 & 7.900e+01 & 1.477e-06 & 4.984e+03 & 1.090e-11 &  \nodata & \nodata \\
2.185e-02 & 9.479e-02 & 1.379e+00 & 5.249e-02 & 8.699e+01 & 1.103e-06 & 5.489e+03 & 8.310e-12 &  \nodata & \nodata \\
2.406e-02 & 1.024e-01 & 1.518e+00 & 4.411e-02 & 9.580e+01 & 8.243e-07 & 6.044e+03 & 6.337e-12 &  \nodata & \nodata \\
2.650e-02 & 1.106e-01 & 1.672e+00 & 3.691e-02 & 1.055e+02 & 6.162e-07 & 6.656e+03 & 4.833e-12 &  \nodata & \nodata \\
2.918e-02 & 1.194e-01 & 1.841e+00 & 3.073e-02 & 1.162e+02 & 4.610e-07 & 7.329e+03 & 3.685e-12 &  \nodata & \nodata \\
3.213e-02 & 1.286e-01 & 2.027e+00 & 2.547e-02 & 1.279e+02 & 3.453e-07 & 8.071e+03 & 2.810e-12 &  \nodata & \nodata \\
3.539e-02 & 1.384e-01 & 2.233e+00 & 2.103e-02 & 1.409e+02 & 2.590e-07 & 8.887e+03 & 2.142e-12 &  \nodata & \nodata \\
	\enddata
    \tablecomments{Differential Intensity units: (m$^2$ s sr GeV/n)$^{-1}$.}
\end{deluxetable}

\begin{deluxetable}{cccccccccc}%[p!]
	\tabletypesize{\footnotesize}
	\tablecolumns{10}
	\tablewidth{0mm}
	\tablecaption{$Z=9$ -- total fluorine LIS with the primary component\label{Tbl-FluorineLIS-Rigi_primes}}
	\tablehead{
		\colhead{Rigidity} & \colhead{Differential} &
		\colhead{Rigidity} & \colhead{Differential} &
		\colhead{Rigidity} & \colhead{Differential} &
		\colhead{Rigidity} & \colhead{Differential} &
		\colhead{Rigidity} & \colhead{Differential} \\ [-2ex] 
		\colhead{GV} & \colhead{intensity} &
		\colhead{GV} & \colhead{intensity} &
		\colhead{GV} & \colhead{intensity} &
		\colhead{GV} & \colhead{intensity} &
		\colhead{GV} & \colhead{intensity}  
	}
	\startdata
9.114e-02 & 1.774e-04 & 7.359e-01 & 3.673e-02 & 1.018e+01 & 2.886e-03 & 5.322e+02 & 2.275e-08 & 3.346e+04 & 1.991e-13\\
9.565e-02 & 2.032e-04 & 7.735e-01 & 4.051e-02 & 1.105e+01 & 2.312e-03 & 5.858e+02 & 1.727e-08 & 3.684e+04 & 1.518e-13\\
1.004e-01 & 2.306e-04 & 8.132e-01 & 4.449e-02 & 1.200e+01 & 1.845e-03 & 6.449e+02 & 1.312e-08 & 4.057e+04 & 1.157e-13\\
1.053e-01 & 2.619e-04 & 8.551e-01 & 4.866e-02 & 1.304e+01 & 1.467e-03 & 7.100e+02 & 9.985e-09 & 4.467e+04 & 8.816e-14\\
1.105e-01 & 2.974e-04 & 8.992e-01 & 5.297e-02 & 1.419e+01 & 1.161e-03 & 7.816e+02 & 7.603e-09 & 4.919e+04 & 6.718e-14\\
1.160e-01 & 3.378e-04 & 9.459e-01 & 5.735e-02 & 1.545e+01 & 9.153e-04 & 8.605e+02 & 5.794e-09 & 5.417e+04 & 5.120e-14\\
1.217e-01 & 3.838e-04 & 9.952e-01 & 6.177e-02 & 1.684e+01 & 7.185e-04 & 9.474e+02 & 4.417e-09 & 5.965e+04 & 3.901e-14\\
1.277e-01 & 4.361e-04 & 1.047e+00 & 6.616e-02 & 1.836e+01 & 5.616e-04 & 1.043e+03 & 3.369e-09 & 6.569e+04 & 2.973e-14\\
1.341e-01 & 4.957e-04 & 1.103e+00 & 7.046e-02 & 2.004e+01 & 4.371e-04 & 1.148e+03 & 2.571e-09 & 7.233e+04 & 2.265e-14\\
1.407e-01 & 5.634e-04 & 1.161e+00 & 7.458e-02 & 2.189e+01 & 3.389e-04 & 1.264e+03 & 1.963e-09 & 7.965e+04 & 1.726e-14\\
1.476e-01 & 6.404e-04 & 1.223e+00 & 7.844e-02 & 2.392e+01 & 2.618e-04 & 1.392e+03 & 1.499e-09 & 8.771e+04 & 1.315e-14\\
1.549e-01 & 7.281e-04 & 1.289e+00 & 8.193e-02 & 2.616e+01 & 2.014e-04 & 1.533e+03 & 1.144e-09 & 9.658e+04 & 1.002e-14\\
1.626e-01 & 8.278e-04 & 1.358e+00 & 8.499e-02 & 2.862e+01 & 1.544e-04 & 1.688e+03 & 8.738e-10 & 1.064e+05 & 7.638e-15\\
1.707e-01 & 9.412e-04 & 1.432e+00 & 8.748e-02 & 3.133e+01 & 1.181e-04 & 1.858e+03 & 6.672e-10 & 1.171e+05 & 5.820e-15\\
1.791e-01 & 1.070e-03 & 1.511e+00 & 8.934e-02 & 3.431e+01 & 9.005e-05 & 2.046e+03 & 5.094e-10 & 1.290e+05 & 4.434e-15\\
1.880e-01 & 1.217e-03 & 1.595e+00 & 9.050e-02 & 3.759e+01 & 6.851e-05 & 2.253e+03 & 3.891e-10 & 1.420e+05 & 3.379e-15\\
1.973e-01 & 1.384e-03 & 1.684e+00 & 9.087e-02 & 4.121e+01 & 5.201e-05 & 2.481e+03 & 2.972e-10 & 1.564e+05 & 2.575e-15\\
2.070e-01 & 1.574e-03 & 1.780e+00 & 9.037e-02 & 4.518e+01 & 3.941e-05 & 2.731e+03 & 2.270e-10 & 1.722e+05 & 1.962e-15\\
2.173e-01 & 1.789e-03 & 1.881e+00 & 8.898e-02 & 4.957e+01 & 2.979e-05 & 3.008e+03 & 1.734e-10 & 1.896e+05 & 1.495e-15\\
2.280e-01 & 2.034e-03 & 1.990e+00 & 8.671e-02 & 5.439e+01 & 2.249e-05 & 3.312e+03 & 1.324e-10 & 2.088e+05 & 1.139e-15\\
2.393e-01 & 2.312e-03 & 2.107e+00 & 8.361e-02 & 5.970e+01 & 1.695e-05 & 3.647e+03 & 1.011e-10 &  \nodata & \nodata \\
2.512e-01 & 2.628e-03 & 2.232e+00 & 7.975e-02 & 6.555e+01 & 1.275e-05 & 4.015e+03 & 7.718e-11 &  \nodata & \nodata \\
2.637e-01 & 2.987e-03 & 2.366e+00 & 7.534e-02 & 7.199e+01 & 9.585e-06 & 4.421e+03 & 5.892e-11 &  \nodata & \nodata \\
2.767e-01 & 3.394e-03 & 2.511e+00 & 7.047e-02 & 7.908e+01 & 7.196e-06 & 4.869e+03 & 4.497e-11 &  \nodata & \nodata \\
2.905e-01 & 3.855e-03 & 2.666e+00 & 6.520e-02 & 8.688e+01 & 5.397e-06 & 5.361e+03 & 3.433e-11 &  \nodata & \nodata \\
3.049e-01 & 4.377e-03 & 2.833e+00 & 5.966e-02 & 9.548e+01 & 4.044e-06 & 5.903e+03 & 2.620e-11 &  \nodata & \nodata \\
3.201e-01 & 4.968e-03 & 3.014e+00 & 5.401e-02 & 1.049e+02 & 3.027e-06 & 6.501e+03 & 1.999e-11 &  \nodata & \nodata \\
3.360e-01 & 5.636e-03 & 3.210e+00 & 4.839e-02 & 1.154e+02 & 2.264e-06 & 7.158e+03 & 1.525e-11 &  \nodata & \nodata \\
3.527e-01 & 6.390e-03 & 3.421e+00 & 4.294e-02 & 1.268e+02 & 1.692e-06 & 7.882e+03 & 1.164e-11 &  \nodata & \nodata \\
3.702e-01 & 7.242e-03 & 3.650e+00 & 3.774e-02 & 1.395e+02 & 1.264e-06 & 8.679e+03 & 8.876e-12 &  \nodata & \nodata \\
3.887e-01 & 8.202e-03 & 3.899e+00 & 3.283e-02 & 1.534e+02 & 9.437e-07 & 9.558e+03 & 6.771e-12 &  \nodata & \nodata \\
4.081e-01 & 9.284e-03 & 4.169e+00 & 2.830e-02 & 1.687e+02 & 7.047e-07 & 1.052e+04 & 5.164e-12 &  \nodata & \nodata \\
4.285e-01 & 1.050e-02 & 4.463e+00 & 2.417e-02 & 1.856e+02 & 5.263e-07 & 1.159e+04 & 3.938e-12 &  \nodata & \nodata \\
4.499e-01 & 1.186e-02 & 4.783e+00 & 2.048e-02 & 2.042e+02 & 3.931e-07 & 1.276e+04 & 3.003e-12 &  \nodata & \nodata \\
4.724e-01 & 1.339e-02 & 5.132e+00 & 1.726e-02 & 2.247e+02 & 2.938e-07 & 1.405e+04 & 2.290e-12 &  \nodata & \nodata \\
4.961e-01 & 1.510e-02 & 5.513e+00 & 1.445e-02 & 2.472e+02 & 2.198e-07 & 1.547e+04 & 1.746e-12 &  \nodata & \nodata \\
5.210e-01 & 1.700e-02 & 5.929e+00 & 1.203e-02 & 2.720e+02 & 1.646e-07 & 1.704e+04 & 1.331e-12 &  \nodata & \nodata \\
5.472e-01 & 1.911e-02 & 6.384e+00 & 9.968e-03 & 2.993e+02 & 1.234e-07 & 1.876e+04 & 1.015e-12 &  \nodata & \nodata \\
	\enddata
	\tablecomments{Differential Intensity units: (m$^2$ s sr GV)$^{-1}$.}
\end{deluxetable}

\begin{deluxetable}{cccccccccc}%[p!]
	\tabletypesize{\footnotesize}
	\tablecolumns{10}
	\tablewidth{0mm}
	\tablecaption{$Z=9$ -- total fluorine LIS with the primary component\label{Tbl-FluorineLIS-EKin_primes}}
	\tablehead{
		\colhead{$E_{\rm kin}$} & \colhead{Differential} &
		\colhead{$E_{\rm kin}$} & \colhead{Differential} &
		\colhead{$E_{\rm kin}$} & \colhead{Differential} &
		\colhead{$E_{\rm kin}$} & \colhead{Differential} &
		\colhead{$E_{\rm kin}$} & \colhead{Differential} \\ [-2ex] 
		\colhead{GeV/n} & \colhead{intensity} &
		\colhead{GeV/n} & \colhead{intensity} &
		\colhead{GeV/n} & \colhead{intensity} &
		\colhead{GeV/n} & \colhead{intensity} &
		\colhead{GeV/n} & \colhead{intensity}  
	}
	\startdata
1.000e-03 & 8.088e-03 & 6.309e-02 & 2.212e-01 & 3.981e+00 & 6.206e-03 & 2.512e+02 & 4.803e-08 & 1.585e+04 & 4.204e-13\\
1.101e-03 & 8.828e-03 & 6.948e-02 & 2.336e-01 & 4.384e+00 & 4.957e-03 & 2.766e+02 & 3.646e-08 & 1.745e+04 & 3.204e-13\\
1.213e-03 & 9.552e-03 & 7.651e-02 & 2.458e-01 & 4.827e+00 & 3.947e-03 & 3.046e+02 & 2.771e-08 & 1.922e+04 & 2.442e-13\\
1.335e-03 & 1.034e-02 & 8.425e-02 & 2.576e-01 & 5.315e+00 & 3.131e-03 & 3.354e+02 & 2.108e-08 & 2.116e+04 & 1.861e-13\\
1.470e-03 & 1.119e-02 & 9.277e-02 & 2.689e-01 & 5.853e+00 & 2.475e-03 & 3.693e+02 & 1.605e-08 & 2.330e+04 & 1.418e-13\\
1.619e-03 & 1.211e-02 & 1.022e-01 & 2.793e-01 & 6.446e+00 & 1.948e-03 & 4.067e+02 & 1.223e-08 & 2.566e+04 & 1.081e-13\\
1.783e-03 & 1.311e-02 & 1.125e-01 & 2.888e-01 & 7.098e+00 & 1.527e-03 & 4.478e+02 & 9.325e-09 & 2.825e+04 & 8.236e-14\\
1.963e-03 & 1.420e-02 & 1.239e-01 & 2.971e-01 & 7.816e+00 & 1.192e-03 & 4.931e+02 & 7.112e-09 & 3.111e+04 & 6.276e-14\\
2.162e-03 & 1.538e-02 & 1.364e-01 & 3.042e-01 & 8.607e+00 & 9.272e-04 & 5.430e+02 & 5.428e-09 & 3.426e+04 & 4.783e-14\\
2.381e-03 & 1.667e-02 & 1.502e-01 & 3.097e-01 & 9.478e+00 & 7.184e-04 & 5.980e+02 & 4.144e-09 & 3.773e+04 & 3.644e-14\\
2.622e-03 & 1.806e-02 & 1.654e-01 & 3.136e-01 & 1.044e+01 & 5.545e-04 & 6.585e+02 & 3.164e-09 & 4.155e+04 & 2.777e-14\\
2.887e-03 & 1.957e-02 & 1.822e-01 & 3.156e-01 & 1.149e+01 & 4.264e-04 & 7.251e+02 & 2.416e-09 & 4.575e+04 & 2.116e-14\\
3.179e-03 & 2.121e-02 & 2.006e-01 & 3.157e-01 & 1.266e+01 & 3.268e-04 & 7.985e+02 & 1.845e-09 & 5.038e+04 & 1.612e-14\\
3.501e-03 & 2.298e-02 & 2.209e-01 & 3.137e-01 & 1.394e+01 & 2.498e-04 & 8.793e+02 & 1.409e-09 & 5.548e+04 & 1.229e-14\\
3.855e-03 & 2.491e-02 & 2.432e-01 & 3.095e-01 & 1.535e+01 & 1.904e-04 & 9.682e+02 & 1.075e-09 & 6.109e+04 & 9.362e-15\\
4.245e-03 & 2.700e-02 & 2.678e-01 & 3.033e-01 & 1.690e+01 & 1.448e-04 & 1.066e+03 & 8.214e-10 & 6.727e+04 & 7.133e-15\\
4.675e-03 & 2.927e-02 & 2.949e-01 & 2.949e-01 & 1.861e+01 & 1.099e-04 & 1.174e+03 & 6.274e-10 & 7.408e+04 & 5.435e-15\\
5.148e-03 & 3.173e-02 & 3.248e-01 & 2.843e-01 & 2.049e+01 & 8.327e-05 & 1.293e+03 & 4.792e-10 & 8.157e+04 & 4.141e-15\\
5.669e-03 & 3.439e-02 & 3.577e-01 & 2.717e-01 & 2.257e+01 & 6.295e-05 & 1.424e+03 & 3.660e-10 & 8.983e+04 & 3.156e-15\\
6.242e-03 & 3.728e-02 & 3.938e-01 & 2.573e-01 & 2.485e+01 & 4.750e-05 & 1.568e+03 & 2.795e-10 & 9.892e+04 & 2.404e-15\\
6.874e-03 & 4.041e-02 & 4.337e-01 & 2.414e-01 & 2.736e+01 & 3.579e-05 & 1.726e+03 & 2.134e-10 &  \nodata & \nodata \\
7.569e-03 & 4.379e-02 & 4.776e-01 & 2.244e-01 & 3.013e+01 & 2.693e-05 & 1.901e+03 & 1.629e-10 &  \nodata & \nodata \\
8.335e-03 & 4.745e-02 & 5.259e-01 & 2.068e-01 & 3.318e+01 & 2.024e-05 & 2.093e+03 & 1.244e-10 &  \nodata & \nodata \\
9.179e-03 & 5.141e-02 & 5.791e-01 & 1.890e-01 & 3.654e+01 & 1.520e-05 & 2.305e+03 & 9.495e-11 &  \nodata & \nodata \\
1.011e-02 & 5.569e-02 & 6.377e-01 & 1.711e-01 & 4.023e+01 & 1.140e-05 & 2.539e+03 & 7.247e-11 &  \nodata & \nodata \\
1.113e-02 & 6.030e-02 & 7.022e-01 & 1.533e-01 & 4.431e+01 & 8.538e-06 & 2.795e+03 & 5.530e-11 &  \nodata & \nodata \\
1.226e-02 & 6.528e-02 & 7.733e-01 & 1.361e-01 & 4.879e+01 & 6.391e-06 & 3.078e+03 & 4.220e-11 &  \nodata & \nodata \\
1.350e-02 & 7.065e-02 & 8.515e-01 & 1.198e-01 & 5.373e+01 & 4.780e-06 & 3.390e+03 & 3.220e-11 &  \nodata & \nodata \\
1.486e-02 & 7.642e-02 & 9.377e-01 & 1.046e-01 & 5.916e+01 & 3.572e-06 & 3.733e+03 & 2.456e-11 &  \nodata & \nodata \\
1.637e-02 & 8.263e-02 & 1.033e+00 & 9.049e-02 & 6.515e+01 & 2.668e-06 & 4.110e+03 & 1.874e-11 &  \nodata & \nodata \\
1.802e-02 & 8.930e-02 & 1.137e+00 & 7.763e-02 & 7.174e+01 & 1.993e-06 & 4.526e+03 & 1.429e-11 &  \nodata & \nodata \\
1.985e-02 & 9.646e-02 & 1.252e+00 & 6.605e-02 & 7.900e+01 & 1.488e-06 & 4.984e+03 & 1.090e-11 &  \nodata & \nodata \\
2.185e-02 & 1.041e-01 & 1.379e+00 & 5.575e-02 & 8.699e+01 & 1.111e-06 & 5.489e+03 & 8.315e-12 &  \nodata & \nodata \\
2.406e-02 & 1.123e-01 & 1.518e+00 & 4.675e-02 & 9.580e+01 & 8.300e-07 & 6.044e+03 & 6.341e-12 &  \nodata & \nodata \\
2.650e-02 & 1.210e-01 & 1.672e+00 & 3.901e-02 & 1.055e+02 & 6.203e-07 & 6.656e+03 & 4.835e-12 &  \nodata & \nodata \\
2.918e-02 & 1.303e-01 & 1.841e+00 & 3.239e-02 & 1.162e+02 & 4.640e-07 & 7.329e+03 & 3.687e-12 &  \nodata & \nodata \\
3.213e-02 & 1.402e-01 & 2.027e+00 & 2.676e-02 & 1.279e+02 & 3.474e-07 & 8.071e+03 & 2.811e-12 &  \nodata & \nodata \\
3.539e-02 & 1.505e-01 & 2.233e+00 & 2.202e-02 & 1.409e+02 & 2.605e-07 & 8.887e+03 & 2.143e-12 &  \nodata & \nodata \\
	\enddata
	\tablecomments{Differential Intensity units: (m$^2$ s sr GeV/n)$^{-1}$.}
\end{deluxetable}

\end{document}